\documentclass{jfm}
\usepackage{graphicx}
\usepackage{epstopdf}
\usepackage{newtxtext}
\usepackage{newtxmath}
\usepackage{natbib}
\usepackage{tikz}
\usetikzlibrary{arrows,decorations.markings}
\usetikzlibrary{plotmarks}
\usetikzlibrary{positioning}
\usepackage{pgfplots}
\usepgfplotslibrary{groupplots}
\usepackage{subcaption}
\usetikzlibrary{calc}
\usepackage[ruled,vlined]{algorithm2e}
\usepgfplotslibrary{groupplots}
\usepackage{multirow}
\usepackage{comment}
\usepackage{hyperref}
\hypersetup{
    colorlinks = true,
    urlcolor   = blue,
    citecolor  = black,
}
\usepackage{soul}

\newcommand{\RomanNumeralCaps}[1]
\linenumbers

\newcommand{\DparD}[2]{\dfrac{\partial #1}{\partial #2}} 

\newcommand{\vu}{ \boldsymbol{u}}
\newcommand{\vup}{\boldsymbol{v}}
\newcommand{\wu}{\boldsymbol{w}}
\newcommand{\wuf}{\boldsymbol{w}_{\boldsymbol{f}}}
\newcommand{\vuf}{ \boldsymbol{u_f} }
 
\newcommand{\vufp}{ \boldsymbol{v_f} }

\definecolor{firstColor}{rgb}{0, 0.4470, 0.7410}
\definecolor{secondColor}{rgb}{0.8500, 0.3250, 0.0980}
\definecolor{thirdColor}{rgb}{0.9290, 0.6940, 0.1250}
\definecolor{fourthColor}{rgb}{0.4660, 0.6740, 0.1880}

\newcommand{\vum}{ \boldsymbol{u_m} } 
 
\newcommand{\vump}{ \boldsymbol{v_m} } 

\newcommand{\vuj}{ \boldsymbol{v_j} } 
 \newcommand{\vufbar}{ \bar{\boldsymbol{u}}_{\boldsymbol{f}}}
  \newcommand{\vumbar}{ \bar{\boldsymbol{u}}_{\boldsymbol{m}}}

\newcommand{\vecf}{ \boldsymbol{f} } 
\newcommand{\vecg}{ \boldsymbol{g} } 
\newcommand{\matA}{ \mathsfbi{A} } 
\newcommand{\matB}{ \mathsfbi{B} } 

\newcommand{\alBJSJ}{\alpha}
\newcommand{\dhat}{\hat{d}}
\newcommand{\dhatc}{\hat{d}^*}
\newcommand{\Ram}{Ra_m}
\newcommand{\Da}{Da}
\newcommand{\Prm}{Pr_{m}}
\newcommand{\Raf}{Ra_{f}}

\newcommand{\Ramc}{\textrm{Ra}_{m,c}}

\newcommand{\Ramcu}{\Ram^*}
\newcommand{\Rafcu}{\Raf^*}

\title{Predicting convection configurations in coupled fluid-porous systems}

\author{Matthew McCurdy\aff{1}
  \corresp{\email{mtmccurdy@owu.edu}},
  Nicholas J. Moore\aff{2},
  \and Xiaoming Wang\aff{3}\aff{4}
 }

\affiliation{\aff{1}Department of Mathematics and Computer Science, Ohio Wesleyan University, Delaware, OH 43015, USA
\aff{2}Department of Mathematics, Colgate University, Hamilton, NY 13346, USA
\aff{3} Department of Mathematics, SUSTech International Center for Mathematics, National Center for Applied Mathematics Shenzhen, Guangdong Provincial Key Laboratory of Computational Science and Material Design, Southern University of Science and Technology, Shenzhen 518055, PR China
\aff{4} Department of of Mathematics and Statistics, Missouri University of Science and Technology, Rolla, MO, 65409 USA
}

\begin{document}
\maketitle

\begin{abstract}
A ubiquitous arrangement in nature is a free-flowing fluid coupled to a porous medium, for example a river or lake lying above a porous bed. Depending on the environmental conditions, thermal convection can occur and may be confined to the clear fluid region, forming shallow convection cells, or it can penetrate into the porous medium, forming deep cells. Here, we combine three complementary approaches --- linear stability analysis, fully nonlinear numerical simulations, and a coarse-grained model --- to determine the circumstances that lead to each configuration. The coarse-grained model yields an explicit formula for the transition between deep and shallow convection in the physically relevant limit of small Darcy number. Near the onset of convection, all three of the approaches agree, validating the predictive capability of the explicit formula. The numerical simulations extend these results into the strongly nonlinear regime, revealing novel hybrid configurations in which the flow exhibits a dynamic shift from shallow to deep convection. This hybrid shallow-to-deep convection begins with small, random initial data, progresses through a metastable shallow state, and arrives at the preferred steady-state of deep convection. We construct a phase diagram that incorporates information from all three approaches and depicts the regions in parameter space that give rise to each convective state.
\end{abstract}



\section{Introduction}
\label{sec:intro}

Convection in fluid-porous systems is a universally-observed phenomenon, with applications arising in technological, geophysical, and astrophysical settings.
In technological and industrial applications, fluid-porous convection is prevalent in heat sinks and cooling technologies found in laptops and computers \citep[see][]{al2017analysis, yu2010natural, yu2011optimum}, as well as in the solidification of alloys \citep[see][]{lebars2006a,lebars2006b}. 
Geophysical examples can be seen in the coupled fluid-porous flow of sub-glacial or dry salt lakes \citep[see][]{couston2021turbulent,hirata2012onset,lasser2021stability}, in carbon-dioxide sequestration or the flow of oil in underground reservoirs \citep[see][]{allen1984,ewing1997,huppert2014fluid}, and in contaminant transport in sub-soil water reservoirs \citep[see][]{curran1990parallel,allen1992solute}. Other geophysical applications that feature natural convection include plate tectonics  \citep[see][]{zhang2000periodic, Mac2018}, cave ventilation  \citep[see][]{Caves2019}, and morphological formation from solute-laden flows \citep[see][]{wykes2018self, mac2020ultra, mac2022morphological}.
The phenomenon of convection also extends far beyond Earth and into astrophysical applications, such as in moons of Saturn and Jupiter \citep[see][]{choblet2017powering,le2020internally,vilella2020tidally}. 

Although flow in fluid-porous systems has been a staple of the research community since Saffman and Jones' work in the 1970s \citep[see][]{saffman1971,jones1973}, convection in these systems still poses unique and timely questions.
Recent years have seen a resurgence of research on fluid-porous convection from a variety of viewpoints, including conducting nonlinear stability analysis, exploring bifurcations, and developing stable numerical schemes for solutions (see \citet[][]{mccurdy2019convection, han2020dynamic, wang2021global,chen2020uniquely,chen2021conservative}.
One recurring theme observed in many of these works is the contrast between {\em deep convection}, in which convection cells occupy the entirety of the coupled domain, and {\em shallow convection}, in which cells only circulate in the free-flow region.

The arrangement we consider is a saturated porous layer lying beneath a clear fluid region that is free from obstructions, illustrated with the schematic in figure~\ref{fig:dom}. 
This entire coupled system is heated from below.
Each domain has a governing set of equations for the fluid flow --- Navier-Stokes for the free-flow region, and Darcy's equations for the porous medium --- with a set of conditions imposed along the interface between the two. 
If the temperature difference between the lower and upper plates rises above a critical threshold, then the conductive state becomes unstable and gives way to natural convection. 
If the temperature difference is sufficiently small, the conductive state remains stable.

In previous work, \citet{mccurdy2019convection} conducted linear and nonlinear stability analyses of the superposed system. With certain parameter regimes, the bimodal marginal stability curves suggested that a small change in the depth ratio of the two regions could trigger a drastic change in the convection patterns. 
The interested reader can look ahead to figure~\ref{fig:flowConfigs}a--b to see the stark contrast between the flow profiles for depth ratios of $\dhat = .18$ and $\dhat = .19$ respectively.
Indeed, slightly altering the depth ratio induces a qualitative shift in flow behavior, as originally observed by \citet{chen1988}.
This drastic change spurred \citet{mccurdy2019convection} to develop a simple, coarse-grained model to narrow down the parameter ranges for which the transition occurs. 
This simple model, which neglected any coupling between the free-flow and porous regions, provided promising results and laid the groundwork for the current study.

Here, we improve upon the model to better account for the flow-conditions at the interface between the fluid and the porous medium. The analysis results in a simple, explicit formula for the critical depth ratio at which shallow convection transitions to deep convection.
We expect the new formula to be relevant for geophysical applications, such as predicting the penetration of tracers into groundwater, and industrial applications, for instance controlling heat dissipation by appropriately choosing the depth and/or porosity of a heat sink.
To test the new model, we conduct numerical simulations of the fully nonlinear, coupled Navier-Stokes-Darcy-Boussinesq system. 
Computing numerical solutions to this system presents several challenges, for example accurately representing the sharp transitions in physical properties (e.g.~density, conductivity, diffusivity) across interfaces and achieving stable time-stepping in face of the nonlinearities present in the Navier-Stokes equations. 
As the numerical scheme is detailed, we explain how we address each of these challenges.
Ultimately, the fully nonlinear numerical simulations provide a more comprehensive picture of fluid-porous convection, revealing novel flow configurations not easily predicted by stability analysis or the coarse-grained model.

A few recent works have focused on convection in superposed fluid-porous layers from a numerical perspective. 
\cite{zhang2020well} used a finite element method (FEM) to study the stationary Navier-Stokes-Darcy-Boussinesq system and investigated the well-posedness of their finite element (FE) approximation. 
Other works have numerically examined convection in related systems \citep[see][]{le2021high,al2017analysis,tatsuo1986numerical, valencia2001study}, albeit with different governing equations (e.g.~the Brinkman system instead of the Darcy system, or the Stokes equations instead of Navier-Stokes). 
Some works have explored schemes to couple the Cahn-Hilliard equations with fluid-porous flow, \citep[see][]{chen2020uniquely,chen2021conservative}, while others have taken an analytical approach to convection in coupled layers. 
For example, \citet[][]{han2020dynamic} recently examined transitions in the same Navier-Stokes-Darcy-Boussinesq system considered here, although through a different lens. Their main focus was the transition from a conductive to a convective state as the Raleigh number increases. The work of \citet[][]{han2020dynamic} rigorously showed that transitions exist between different convective profiles, like deep and shallow convection, and noted how the transitions behaved-- as continuous transitions or jump transitions-- around their critical Rayleigh numbers. In this work, we examine a similar kind of transition as we develop a model to predict the parameter regimes where the switch between deep and shallow convection takes place. This kind of transition, dubbed a `dynamic transition' by \citet[][]{han2020dynamic}, is a bifurcation of the system as one moves through the parameter space and has been observed in a number of papers, \citep[][]{chen1988,chen1989experimental,chen1992convection,mckay1998onset,straughan2002a,hirata2007linear,yin2013stability,mccurdy2019convection}. Our model improves the prediction of our previous model by utilizing an {\em open boundary condition} at the interface when calculating the critical Rayleigh numbers in our theory.  Additionally, we note a second kind of transition in this manuscript. This transition, which we refer to as a `dynamic shift,' is associated with the time evolution of the system where the conductive state transitions through a metastable shallow-convection state en route to its steady state of deep convection. Our numerical simulations shed light on this new flow configuration, shallow-to-deep convection.

The article is organized as follows. 
In section \ref{sec:system}, we present the system of equations, interface conditions, and the nondimensionalized system. 
Then, the transition theory-- one of the two main contributions of this work-- is introduced in section \ref{sec:theory} along with results showing its efficacy.
Next, we detail our numerical scheme using a FEM in section \ref{sec:numerics}, and we note how the treatment of interfacial and nonlinear terms allows us to write the system as a set of linear, sequentially decoupled equations. 
Results are shown in section \ref{sec:results} along with a discussion of how our three complementary approaches-- stability analysis, fully nonlinear numerical simulations, and the coarse-grained model-- agree to provide a more complete picture of convection in fluid-porous systems. Our results showcase the second main contribution of this paper: the novel convection pattern of shallow-to-deep convection.

\section{The coupled Navier-Stokes-Darcy System}
\label{sec:system}
In this section, we present the governing equations, detail the boundary and interface conditions, and introduce the nondimensional system.

\subsection{Governing equations}
In the free-flow zone, we use the same system studied by \cite{mccurdy2019convection,han2020dynamic}~-- the incompressible Navier-Stokes equations with constant viscosity and the Boussinesq approximation, coupled with the advection-diffusion equations for heat:
\begin{equation}
\left. \begin{array}{rcl}
   \rho_0\left( \DparD{\vuf}{t_f} + \left(\vuf \bcdot \bnabla\right) \vuf \right) &=& \bnabla \bcdot\mathsfbi{T}\left(\vuf, p_f \right) - g\rho_0 \left[1-\beta\left(T_f-T_0 \right) \right]\boldsymbol{k}\, , \\[14pt]
   \bnabla \bcdot \vuf &=&0\, , \\[6pt]
       \DparD{T_f}{t_f} + \vuf \bcdot \bnabla T_f &=&\dfrac{\kappa_f}{ \left(\rho_0c_p\right)_f} \bnabla^2T_f\, \, ,
 \end{array}\right\}
  \label{eq:NavStokesSys}
\end{equation}
where $\vuf = (u_f, v_f, w_f)$, $p_f$, and $T_f$ are the free flow velocity, pressure, and temperature, respectively, with $g$, $\rho_0$, $\beta$, and $T_0$ as acceleration due to gravity, the reference density of the fluid, the coefficient of thermal expansion, and the temperature of the conductive state at the interface, respectively.
The stress tensor and rate of strain tensor are defined as $\mathsfbi{T}(\vuf, p_f)=2\mu_0\mathsfbi{D}(\vuf)-p_f\mathsfbi{I}$ and $\mathsfbi{D}(\vuf)=\frac{1}{2}\left(\bnabla\vuf + \bnabla \vuf^{T}\right)$, respectively, with $\mu_0$ as dynamic viscosity and $\boldsymbol{k}$ as the upward pointing unit normal.
Additionally, $\kappa_f$, $c_p$, and $\lambda_f = \kappa_f/\left(\rho_0c_p\right)_f$ are the thermal conductivity of the fluid, specific heat capacity of the fluid, and thermal diffusivity of the fluid, respectively.

For fluid flow in the porous medium, we assume the medium has a small porosity, as is generally applicable to geophysical systems \citep[][]{bear1972,nield2017}. We therefore employ the Darcy-Boussinesq system with the advection-diffusion equation for heat:
\begin{equation}
\left. \begin{array}{rcl}
  \dfrac{\rho_0}{\chi}\DparD{\vum}{t_m} + \dfrac{\mu_0}{\Pi} \vum &=& -\bnabla p_m - g\rho_0 \left[1-\beta\left(T_m-T_L \right) \right] \boldsymbol{k}\, ,\\[14pt]
\bnabla \bcdot \vum &=&0\\[6pt]
  \dfrac{\left(\rho_0c_p\right)_m}{ \left(\rho_0c_p\right)_f}\DparD{T_m}{t_m} + \vum \bcdot \bnabla T_m &=&\dfrac{\kappa_m}{ \left(\rho_0c_p\right)_f} \bnabla^2T_m\, ,
 \end{array}\right\}
  \label{eq:DarcySys}
\end{equation}
where $\vum = (u_m, v_m, w_m),$ $p_m,$ and $T_m$ are the velocity, pressure, and temperature in the porous medium respectively, $\chi$ and $\Pi$ are the porosity and permeability, $\lambda_m = \kappa_m/\left(\rho_0 c_p\right)_f$ is the thermal diffusivity of the medium, and $T_L$ is the temperature at the lower boundary of the domain.
We assume the medium to be homogeneous and isotropic so that the permeability $\Pi$ is constant and scalar-valued.
The thermal conductivity $\kappa_m$ and specific heat capacity $(\rho_0\,c_p)_m$ of the porous medium are defined as averages of the fluid and solid components.
While many studies neglect the time derivative $\partial_t \vum$ in the first equation of \eqref{eq:DarcySys}, retaining this term is useful for energy analysis of the system (see \citet{mccurdy2019convection}).

\begin{figure} 
\centering
\begin{tikzpicture}
\draw[black, line width=1.75pt] (2,1.5) -- (8,1.5);
\draw[black, line width=1.0pt,dashed] (2,0) -- (8,0);
\draw[black, line width=1.75pt] (2,-1.5) -- (8,-1.5);
\node[fill=white,right] 
	at (.5,0) {$z=0$};
\node[fill=white,right] 
	at (.5,1.5) {$z=d_f$};
\node[fill=white,right] 
	at (.5,-1.5) {$z=-d_m$};
\node[fill=white,right] 
	at (3.25, 1.9) {temperature: $T_U$};
\node[fill=white,right] 
	at (3.25, -1.9) {temperature: $T_L$};
\node[fill=white,right] 
	at (3.25, .75) {free-flow, $\Omega_f$};
\node[fill=white,right] 
	at (3.25, -.75) {porous medium, $\Omega_m$};
\node[fill=white] 
	at (7.0, .4) {interface, $\Gamma_i$};
\end{tikzpicture} 
\caption{Schematic of the domain $\Omega = \{(x,y)\in\mathbb{R}^2 \times z\in(-d_m,d_f)\}$, comprised of a free-flow region $\Omega_f$ and a porous medium $\Omega_m$. The two subdomains meet at an interface $\Gamma_i$. The upper and lower boundaries are impermeable and held at constant temperatures $T_U$ and $T_L$, respectively, with $T_L>T_U$. We assume periodicity of the velocity and temperature in the horizontal direction(s) as well.\label{fig:dom}}
\end{figure}
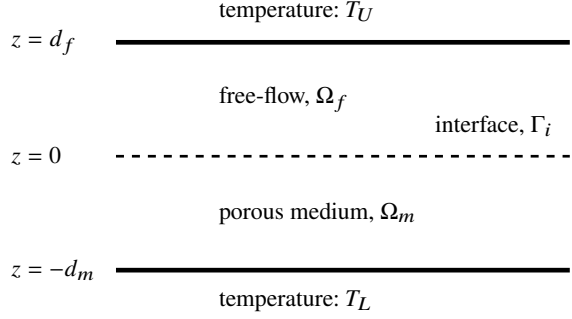

\subsection{Boundary and interface conditions}
The domain, shown in figure~\ref{fig:dom}, consists of flat, horizontal, non-penetrable plates at the top and bottom with a non-deforming interface between the two regions, $\Omega_f=\{(x,y,z)\in \mathbb{R}^2 \times (0,d_f)\}$ for the free flow and $\Omega_m=\{(x,y,z)\in \mathbb{R}^2 \times (-d_m,0)\}$ for the porous medium. 
The temperature is held constant at the top and bottom plates. 
The flow satisfies a free-slip condition at the top and an impermeable condition at the bottom:
\begin{equation}
\left. \begin{array}{ccrcccl}
 T_f = T_U\,, &\quad &\vuf \bcdot \boldsymbol{n} = \frac{\p \vuf_\tau}{\p \boldsymbol{n}}&=0\, &
  \quad &\mbox{on\ }&\quad z=d_f,\\[8pt]
 T_m = T_L\,, &\quad &\vum \bcdot \boldsymbol{n} &=0\, &
  \quad &\mbox{on\ }&\quad z=-d_m,\\[8pt]
 \end{array}\right\}
  \label{eq:BC}
\end{equation}
where ${\vuf}_\tau=(u_f, v_f)$ denotes the tangential (horizontal) components of the velocity at the top of the domain with $\boldsymbol{n}$ as the unit normal vector.

At the interface $\Gamma_i$ $(z=0)$, we require continuity of temperature, heat flux, and the normal component of velocity:
\begin{equation}
\left. \begin{array}{c}
T_f = T_m\, , \\[8pt]
  \kappa_f\,\bnabla T_f \bcdot \boldsymbol{n} =  \kappa_m\,\bnabla T_m \bcdot \boldsymbol{n}\, ,\\[8pt]
  \vuf \bcdot \boldsymbol{n} =  \vum \bcdot \boldsymbol{n}\, .
\end{array} \right\}
\end{equation}
For the last two conditions, we use the Beavers-Joseph-Saffman-Jones (BJSJ) condition \cite{saffman1971} and the Lions interface condition to specify the tangential and normal stresses, respectively. The BJSJ condition, also known as the Navier-slip condition, is

    \begin{equation}\label{eq:BJSJcond}
         -\boldsymbol{\tau} \bcdot \mathsfbi{T} \left(\vuf,p_f \right)\boldsymbol{n} 
         = \frac{\mu_0\,\alBJSJ}{\sqrt{\Pi}}\boldsymbol{\tau} \bcdot \vuf \, ,
    \end{equation}
where $\alBJSJ$ is an empirically determined coefficient and $\boldsymbol{\tau}$ denotes the unit tangent vectors. Lastly, the Lions interface condition is
    \begin{equation}
    \label{normal_cond}
    -\boldsymbol{n} \bcdot \mathsfbi{T}\left(\vuf,p_f \right)\boldsymbol{n} +\frac{\rho_0}{2}\left|\vuf \right|^2 = p_m\, .
    \end{equation}
The inclusion of the $\frac{\rho_0}{2}\left|\vuf \right|^2$ term in this interface condition is essential in conducting the nonlinear stability analysis, \citep[see][]{ccecsmeliouglu2008analysis, ccecsmeliouglu2009primal, chidyagwai2007weak, discacciati2009navier, girault2009dg,mccurdy2019convection}.

 \subsection{System of perturbed variables}
 Instead of working with the physical variables, we consider the perturbed variables; that is, we consider the deviation of the velocity, temperature, and pressure profiles from their conductive steady-states. The conductive state, denoted with an overhead bar, is a stationary fluid and a piecewise-linear temperature:
    \begin{align*}
   \vufbar &= \vumbar = 0\, , \\
     \bar{T}_f&= T_0 + z\frac{T_U - T_0 }{d_f}\, ,\\
     \bar{T}_m &= T_0 + z\frac{T_0 - T_L }{d_m} \, .
    \end{align*}
Here, $T_0$ represents the interface temperature of the conductive solution
    \begin{align*}
    T_0 = \frac{\kappa_m\, d_f\, T_L + \kappa_f \,d_m\, T_U}{\kappa_m \,d_f + \kappa_f \,d_m}\, .
    \end{align*}
If $T_U>T_L$, the conductive state is stable, but if $T_L>T_U$, buoyancy can destabilize the system. Throughout this work, we only consider the case where $T_L>T_U$. Additionally, we choose $\bar{p}_f$ and $\bar{p}_m$ to satisfy
    \begin{align*}
    \nabla \bar{p}_f &= -g\rho_0\left(1-\beta\left(\bar{T}_f-T_0\right)\right)\boldsymbol{k}\, , \\
    \nabla \bar{p}_m &= -g\rho_0\left(1-\beta\left(\bar{T}_m-T_L\right)\right)\boldsymbol{k}\, .
    \end{align*}
With the perturbation variables $\vup_j,$ $\theta_j,$ and $\pi_j$ for $j=\{f,m\}$ regions, we perturb the steady-state solutions:
    \begin{align} \label{eq:perturbedSolns}
    \vuf &=\vufbar + \vufp\, , \quad \vum =  \vumbar +\vump\, \nonumber,\\
    T_f &= \bar{T}_f +\theta_f\, , \quad T_m = \bar{T}_m+\theta_m\, ,  \\
    p_f &= \bar{p}_f +\pi_f\, , \quad p_m= \bar{p}_m + \pi_m \, . \nonumber
    \end{align}
    
Substituting \eqref{eq:perturbedSolns} into the original system produces a system for the {\em perturbed} variables:
\begin{equation}
\left. \begin{array}{rcl}
   \rho_0\left( \DparD{\vufp}{t_f} + \left(\vufp \bcdot \bnabla\right) \vufp \right) &=& \bnabla \bcdot\mathsfbi{T}\left(\vufp, \pi_f \right) + g\rho_0 \beta\theta_f\boldsymbol{k}\, , \\[14pt]
   \bnabla \bcdot \vufp &=&0\, , \\[6pt]
       \DparD{\theta_f}{t_f} + \vufp \bcdot \bnabla \theta_f &=&\lambda_f \bnabla^2\theta_f -\vufp\bcdot\boldsymbol{k} \left(\dfrac{T_U-T_0}{d_f}\right) \, \, ,
 \end{array}\right\}
  \label{eq:NavStokesPertSys}
\end{equation}
for $\Omega_f$;
\begin{equation}
\left. \begin{array}{rcl}
  \dfrac{\rho_0}{\chi}\DparD{\vump}{t_m} + \dfrac{\mu_0}{\Pi} \vump &=& -\bnabla \pi_m + g\rho_0 \beta\theta_m\boldsymbol{k}\, ,\\[14pt]
\bnabla \bcdot \vump &=&0\\[6pt]
  \dfrac{\left(\rho_0c_p\right)_m}{ \left(\rho_0c_p\right)_f}\DparD{\theta_m}{t_m} + \vump \bcdot \bnabla \theta_m &=&\lambda_m \bnabla^2\theta_m -\vump\bcdot\boldsymbol{k} \left(\dfrac{T_0-T_L}{d_m}\right)\, ,
 \end{array}\right\}
  \label{eq:DarcyPertSys}
\end{equation}
for $\Omega_m$; 
  \begin{equation}
\left. \begin{array}{c}
\theta_f = \theta_m\, , \\[8pt]
  \kappa_f\,\bnabla \theta_f \bcdot \boldsymbol{n} =  \kappa_m\,\bnabla \theta_m \bcdot \boldsymbol{n}\, ,\\[8pt]
  \vufp \bcdot \boldsymbol{n} =  \vump \bcdot \boldsymbol{n}\,, \\[8pt]
        -\boldsymbol{\tau} \bcdot \mathsfbi{T} \left(\vufp,\pi_f \right)\boldsymbol{n} 
         = \dfrac{\mu_0\,\alBJSJ}{\sqrt{\Pi}}\boldsymbol{\tau} \bcdot \vufp \, ,\\[8pt]
           -\boldsymbol{n} \bcdot \mathsfbi{T}\left(\vufp,\pi_f \right)\boldsymbol{n} +\dfrac{\rho_0}{2}\left|\vufp \right|^2 = \pi_m\,,
\end{array} \right\}
\end{equation}
 for the interface conditions on $\Gamma_i$; and \begin{equation}
\left. \begin{array}{ccrcccl}
 \theta_f = 0\,, &\quad &\vufp \bcdot \boldsymbol{n} = \frac{\p \vufp_\tau}{\p \boldsymbol{n}}&=0\, &
  \quad &\mbox{on\ }&\quad z=d_f,\\[8pt]
 \theta_m = 0\,, &\quad &\vump \bcdot \boldsymbol{n} &=0\, &
  \quad &\mbox{on\ }&\quad z=-d_m,\\[8pt]
 \end{array}\right\}
  \label{eq:BCPert}
\end{equation}
for the boundary conditions.

\subsection{Nondimensional system}
As in previous work, we nondimensionalize the system using the porous values as a reference \citep[][]{chen1988, mccurdy2019convection, straughan2002a}, with nondimensional variables denoted by tildes:
    \begin{alignat}{5}
    \label{ScalesEqn}
    & \vup_{\boldsymbol{j}} = \tilde{\vup}_{\boldsymbol{j}} \frac{\nu}{d_m}\, , \quad && \boldsymbol{x}_{\boldsymbol{j}} = \tilde{\boldsymbol{x}}_{\boldsymbol{j}} \,d_m\, , \quad    && t_j = \tilde{t}\, \frac{d_m^2}{\lambda_m}\, , \quad && \theta_j = \tilde{\theta}_j \,\frac{\left(T_0-T_L\right)\nu}{\lambda_m}\, ,    \quad    && \pi_j = \tilde{\pi}_j \,\frac{\rho_0\, \nu^2}{d_m^2}\, ,
    \end{alignat} 
    for $j=\{f,m\}$, where $\nu = \mu_0/\rho_0$ is the kinematic viscosity. We also note that the stress tensor has been altered slightly from when it was first introduced; the nondimensional stress tensor is defined as $\widetilde{\mathsfbi{T}}(\vufp, \pi_f)=2\mathsfbi{D}(\vufp)-\pi_f\mathsfbi{I}$.

Substituting the nondimensional variables into equations \eqref{eq:NavStokesPertSys}--\eqref{eq:BCPert} results in the governing equations (after dropping the tildes) for $\Omega_f = \{(x,y,z) \in \mathbb{R}^2\times (0,\dhat)\}$, where $\dhat$ is the ratio of the free-flow depth to that of the porous medium:    
\begin{equation}
\left. \begin{array}{rcl}
    \dfrac{1}{\Prm}\DparD{\vufp}{t} + \left(\vufp \bcdot \bnabla \right)\vufp &=& \bnabla \bcdot\widetilde{\mathsfbi{T}}\left(\vufp, \pi_f \right) +\dfrac{\Ram}{\Da}\,\theta_f\,\boldsymbol{k} , \\[14pt]
   \bnabla \bcdot \vufp &=&0\, , \\[6pt]
     \DparD{\theta_f}{t} + \Prm\,\vufp \bcdot \bnabla \theta_f &=&\epsilon_T\,\bnabla^2 \theta_f + \dfrac{1}{\epsilon_T}\vufp\bcdot\boldsymbol{k} \, ,
 \end{array}\right\}
  \label{eq:NavStokesSysNondim}
\end{equation}
for $\Omega_m = \{(x,y,z) \in \mathbb{R}^2\times (-1,0)\}$:
\begin{equation}
\left. \begin{array}{rcl}
  \dfrac{\Da}{\Prm\, \chi }\DparD{\vump}{t} + \vump &=&-\Da\bnabla \pi_m +\Ram\,\theta_m\,\boldsymbol{k} ,\\[14pt]
\bnabla \bcdot \vump &=&0\,,\\[6pt]
  \varrho\,  \DparD{\theta_m}{t} + \Prm\,\vump \bcdot \bnabla \theta_m &=&\bnabla^2 \theta_m + \vump\bcdot\boldsymbol{k} \,,
 \end{array}\right\}
  \label{eq:DarcySysNondim}
\end{equation}
\noindent and at $\Gamma_i =  \{(x,y,z) \in \mathbb{R}^2\times (z=0)\}$:
\begin{subequations}
\begin{align}
\theta_f &= \theta_m\, , \label{eq:InterfCondNondim.1}\\[6pt]
    \epsilon_T \bnabla \theta_f\bcdot \boldsymbol{n} &=\bnabla \theta_m\bcdot \boldsymbol{n}\, , \label{eq:InterfCondNondim.2}\\[6pt]
  \vufp \bcdot \boldsymbol{n} &=  \vump \bcdot \boldsymbol{n}\, \label{eq:InterfCondNondim.3}\\[6pt]
    -\boldsymbol{\tau} \bcdot \widetilde{\mathsfbi{T}} \left(\vufp,\pi_f \right)\boldsymbol{n} & =\dfrac{\alBJSJ}{\sqrt{\Da}}\left(\boldsymbol{\tau}\bcdot \vufp \right)\,, \label{eq:InterfCondNondim.4}\\[6pt]
    -\boldsymbol{n} \bcdot \widetilde{\mathsfbi{T}} \left(\vufp,\pi_f \right)\boldsymbol{n} +\frac{1}{2}|\vufp|^2 &=\,\pi_m \, .\label{eq:InterfCondNondim.5}
\end{align}
\end{subequations} 
Boundary conditions at the top and bottom of the domain are
\begin{equation}
\left. \begin{array}{ccrcccl}
 \theta_f = 0\,, &\quad &\vufp \bcdot \boldsymbol{n} = \frac{\p \vufp_\tau}{\p \boldsymbol{n}}&=0\, &
  \quad &\mbox{on\ }&\quad z=\hat{d},\\[8pt]
 \theta_m = 0\,, &\quad &\vump \bcdot \boldsymbol{n} &=0\, &
  \quad &\mbox{on\ }&\quad z=-1.\\[8pt]
 \end{array}\right\}
  \label{eq:BCnondim}
\end{equation} 
The nondimensional numbers $\dhat$, $\Prm$, $\Da$, $\epsilon_T$, and $\varrho$ are defined by 
\begin{equation}
\dhat = \frac{d_f}{d_m}\,,\quad\Prm = \frac{\nu}{\lambda_m}\,,\quad \Da =  \frac{\Pi}{d_m^2}\,,\quad \epsilon_T = \frac{\lambda_f}{\lambda_m}\,,\quad \varrho = \frac{\left(\rho_0c_p\right)_m}{ \left(\rho_0c_p\right)_f}\,,
\end{equation} 
and the Rayleigh numbers of both regions are
\begin{equation} \label{eq:rayleighDefns}
\Ram = \frac{g\beta \left(T_L-T_0 \right) \Da\, d_m^3}{\nu \,\lambda_m}\,, \quad 
 \Raf = \frac{\dhat^4}{\Da\,\epsilon_T^2}\Ram \,.
  \end{equation} 
While the Rayleigh number of the porous medium is used throughout this work, the corresponding Rayleigh number of the free-flow region can be determined via the relationship above.

The depth ratio of the two layers $\dhat$ plays a central role in our study. 
Other dimensionless parameters, like $\Prm\,,\epsilon_T,\,\varrho$, represent values inherent to the fluid and/or porous medium. 
In industrial applications, these cannot easily be altered, or it may be impractical to do so.
However, the depth ratio can more readily be changed; for example, by adding or removing fluid from the free-flow layer, the depth ratio varies.
The coarse-grained model presented in the next section details how changing the depth ratio can significantly affect convection profiles.

 The second parameter of interest for geophysical applications is the Darcy number $\Da$--  especially the small Darcy regime since many materials have small permeability values. 
 Relatively impervious materials, like limestone or granite, can have permeabilities between $10^{-18}$ and $10^{-15}m^2$, while more porous media, like sorted gravel or sand, can have permeabilities between $10^{-10}\lesssim \Pi \lesssim 10^{-7}m^2$, \citep[][]{bear1972,nield2017}. 
 The resulting Darcy numbers in experiments and/or simulations is small (typically between $10^{-10}\leq \Da \leq 10^{-5}$), highlighting the necessity to study the relevant small-Darcy regime in more depth, as done in \cite{lyu2021stokes}.

\section{Coarse-grained model for the transition between deep and shallow convection}
\label{sec:theory}

\begin{figure}
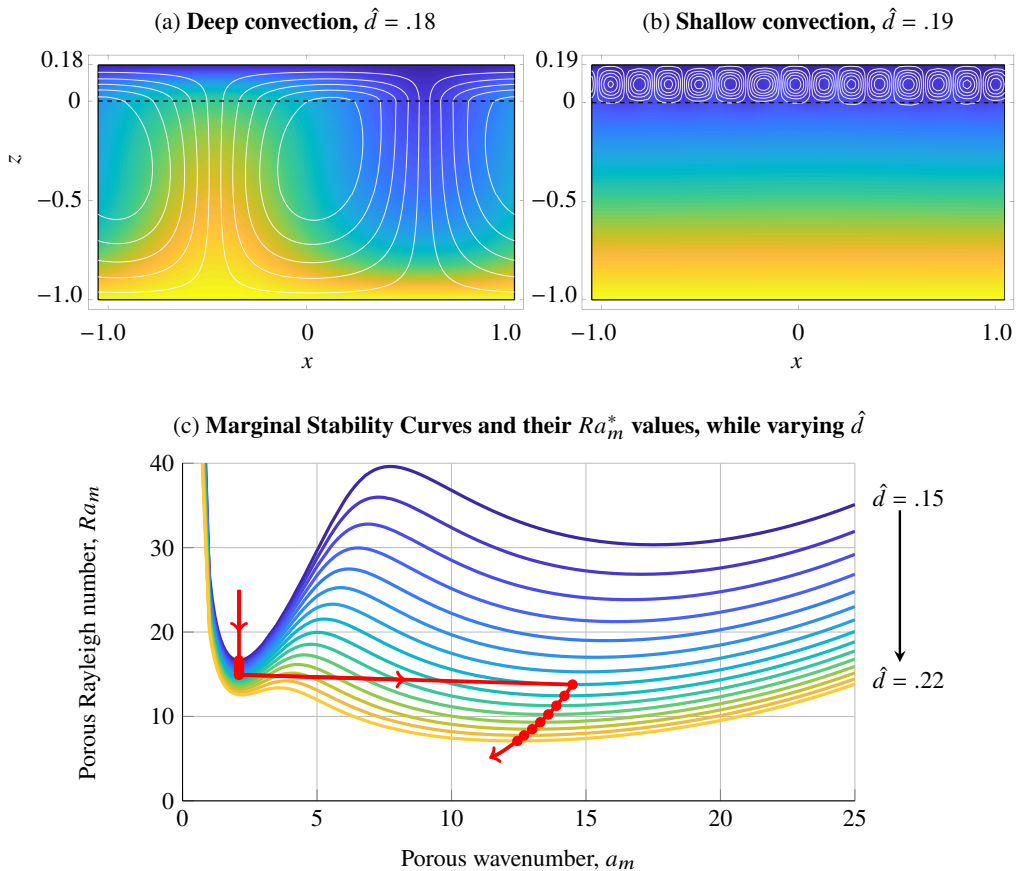
 
\centering
\begin{subfigure}{\linewidth}
\centering
  \input{tikz_files/fig2ab}
\end{subfigure}
\\[\baselineskip \vspace{-.15in}]
\begin{subfigure}{\linewidth}
\centering
\input{tikz_files/fig2c}
\end{subfigure}
\centering
\caption{Marginally stable flow configurations and temperature profiles (color) for (a) $\dhat=.18$ and (b) $\dhat=.19$, with $\sqrt{\Da} = 5.0\times 10^{-3},\, \Prm =.7,\,\epsilon_T=.7,\,\alpha=1.0$ fixed. (c) Marginal stability curves while varying $\dhat$ from $\dhat=.15$ to $\dhat=.22$ by increments of 0.05 with the respective critical Rayleigh numbers $\Ramcu$ shown in red. As $\dhat$ crosses the critical depth ratio of $\dhatc=.181$ for this parameter regime, the most unstable wavenumber jumps from $a_m\approx 2.1$ to $a_m\approx 14.5$. This signifies a sudden transition from deep to shallow convection (at the {\em onset} of convection) as $\dhat$ increases from $\dhat<\dhatc \rightarrow \dhat > \dhatc$.
\label{fig:flowConfigs}}
\end{figure}

\cite{mccurdy2019convection} conducted linear and nonlinear stability analyses of the Navier-Stokes-Darcy-Boussinesq system to determine the threshold Rayleigh number needed for the conductive state to become unstable at a given wavenumber, $a_m$. 
For a fixed depth ratio, figure~\ref{fig:flowConfigs}c shows the collection of these points in the $a_m \text{-} \Ram$ plane, called the {\em marginal stability curve}, which delineates regions of stability from instability.
The critical Rayleigh number $\Ramcu$ is the global minimum of this curve, representing the lowest $\Ram$ value that produces a flow instability. 
The corresponding critical wavenumber $a_m^*$ dictates the flow configuration at the {\em onset} of convection: small $a_m^*$ (i.e.~large wavelength) indicates deep convection cells that occupy the entire coupled domain as seen in figure~\ref{fig:flowConfigs}a, while large $a_m^*$ (i.e.~small wavelength) indicates narrow convection cells that occupy the fluid region only, as seen in figure~\ref{fig:flowConfigs}b.
The bimodal nature of the marginal stability curve can create an abrupt shift between these two flow configurations as the depth ratio changes.

The marginal stability curves shown in figure~\ref{fig:flowConfigs}c are calculated for a range of $\dhat$ values using the linear stability analysis outlined by \cite{mccurdy2019convection}. The global minimum of each curve corresponds to $\Ramcu$ and is indicated by a red dot. The red arrow illustrates the path that connects sequential values of $\Ramcu$ as $\dhat$ increases from 0.15 to 0.22. As seen in the figure, a jump in $a_m^*$ occurs as $\dhat$ changes from 0.18 to 0.19, indicating an abrupt transition from deep convection to shallow convection.

This observation spurred \cite{mccurdy2019convection} to develop a simple model to predict the critical depth ratio $\dhatc$ that distinguishes shallow convection from deep convection. 
A brief outline of this model is as follows. 
First, due to the dimensionless definitions, the depth ratio can be expressed as the following combination of the other parameters:
\begin{equation} 
\label{dhateq}
\hat{d} = \left[\dfrac{\Raf\,\Da\,\epsilon_T^2}{\Ram}\right]^{1/4}\,.
 \end{equation} 
Of particular importance is the appearance of the two Raleigh numbers, $\Raf$ and $\Ram$. 
If the Rayleigh number exceeds its critical value in the fluid region, $\Raf > \Rafcu$, but remains below the critical value in the porous medium, $\Ram < \Ramcu$, then only shallow convection cells would be expected to form. 
On the other hand, if the Rayleigh number exceeds the critical value in both regions, then convection cells can penetrate deeper into the porous domain. 
Hence, the transition between shallow and deep convection should occur when both Rayleigh numbers, $\Raf$ and $\Ram$, become critical simultaneously.
In reality, the critical values, $\Rafcu$ and $\Ramcu$, are coupled to one another through the flow details at the interface. 
However, \cite{mccurdy2019convection} showed that useful estimates could be obtained by treating the fluid and porous regions as {\em uncoupled} for the sake of calculating $\Rafcu$ and $\Ramcu$, and then inputting these values into \eqref{dhateq} to estimate the critical depth ratio $\dhatc$ that defines the transition between shallow and deep convection:
\begin{align} 
\dhatc = \left[\dfrac{\Rafcu}{\Ramcu}\,\Da\,\epsilon_T^2\right]^{1/4}\,.
\end{align}
In particular, \cite{mccurdy2019convection} assumed no-slip and no-penetration conditions along the boundaries of the fluid domain and no-penetration conditions along the boundaries of the porous medium. These assumptions yield  $\Rafcu=1707.8$ and $\Ramcu = 4\pi^2\approx 39.5$, which gives
\begin{align} 
\dhatc = \left[\dfrac{1707.8}{39.5}\,\Da\,\epsilon_T^2\right]^{1/4}\,.
\end{align}
This formula was shown to predict the actual critical depth ratio to within a relative error of 13--17$\%$ for the cases tested by  \cite{mccurdy2019convection}. 
This level of accuracy, while not exceptional, is promising considering that the formula neglects any kind of coupling between the two regions.

Here, we further refine this model by introducing asymptotically weak coupling at the interface \citep[see for example][]{moore2012weak}. 
As demonstrated below, the new model yields improved estimates for $\dhatc$, with relative errors on the order of 0--4\% for sufficiently small $\Da$.
We first asymptotically expand both velocity fields, i.e.~inside the fluid and the medium regions, in small $\Da:=\varepsilon^2$:
\begin{equation}
\vuj^{\varepsilon} = \vuj^{(0)} + \varepsilon\, \vuj^{(1)} + \varepsilon^2 \,\vuj^{(2)} +\hdots,\quad \textrm{for }j\in\{f,m\}\,,
\end{equation}
If $\Da=0$, the Darcy system \eqref{eq:DarcySysNondim} is degenerate, giving a porous-medium velocity that vanishes throughout the domain, $\vump^{(0)} \equiv 0$.
Therefore, interface condition \eqref{eq:InterfCondNondim.3} gives $\vufp^{(0)}\bcdot\boldsymbol{n}=0$ at leading order.
Meanwhile, the leading-order equation of the BJSJ condition \eqref{eq:InterfCondNondim.4} gives $\vufp^{(0)}\bcdot\boldsymbol{\tau}=0$. Thus, both components of the fluid-velocity $\vufp^{(0)}$ vanish at the interface to leading order in $\varepsilon$, and so the no-slip and no-penetration interface conditions assumed by \cite{mccurdy2019convection} are valid to leading-order in the fluid domain. We therefore continue to use the corresponding value $\Rafcu=1707.8$ in the new model.

Approaching the interface from the porous-medium side, though, the condition for an impenetrable boundary, $\vump\bcdot \boldsymbol{n}=0$, is not recovered as the leading-order non-trivial dynamics in small Darcy number. In the improved model, we instead view the top of the porous domain as an {\em open boundary}, along which the pressure is uniform \citep[][]{nield2017}. Due to Darcy's law \eqref{eq:DarcySysNondim}, the condition of constant pressure along a horizontal interface implies that the tangential velocity vanishes $\vump\times \boldsymbol{n}=0$, which produces a critical value of $\Ramcu= 27.1$ at the porous wavenumber of $a_{m,1}^*=2.3$ for the {\em uncoupled} porous medium  \citep[see][]{nield2017,TYVAND200282}. Although we have been so far unable to rigorously justify the use of this condition from first principles, we observe that, in practice, it yields significantly improved estimates for $\Ramcu$ and $a_{m,1}^*$ as $\Da \to 0$. Table \ref{tab:critRam} demonstrates this idea by showing the true critical values $\Ramcu$ for the {\em coupled} system and their critical wavenumbers as calculated numerically by the linear stability analysis, for a range of Darcy numbers and three values of $\epsilon_T$. The table shows that as $\Da \to 0$, $\Ramcu$ is well approximated by 27.1 in each case, with relative errors on the order of 1-2\%.

\begin{table}
  \begin{center}
\def~{\hphantom{0}}
  \begin{tabular}{l | ll | ll | ll }
     & \multicolumn{2}{c|}{\underline{$\epsilon_T=0.5$}}  & \multicolumn{2}{c|}{\underline{$\epsilon_T=0.7$}}                   & \multicolumn{2}{c}{\underline{$\epsilon_T=1.0$}}  \\[3pt]
        $\sqrt{\Da}$ &  \textbf{$a_{m,1}^*$} & \textbf{$\Ramcu$}&  \textbf{$a_{m,1}^*$} & \textbf{$\Ramcu$}&  \textbf{$a_{m,1}^*$} & \textbf{$\Ramcu$} \\[4pt]
        $1.0\times 10^{-2}$   & $2.1$ & $11.04$ & $2.1$ & $11.66$ & $2.1$ & $12.34$    \\ 
        $5.0\times 10^{-3}$   & $2.1$ & $14.05$ & $2.1$ & $14.85$ & $2.1$ & $15.72$    \\ 
        $2.5\times 10^{-3}$   & $2.1$ & $16.75$ & $2.1$ & $17.60$ & $2.1$ & $18.50$    \\ 
        $1.0\times 10^{-3}$   & $2.1$ & $19.76$ & $2.2$ & $20.50$ & $2.2$ & $21.28$    \\ 
        $5.0\times 10^{-4}$   & $2.2$ & $21.55$ & $2.2$ & $22.20$ & $2.2$ & $22.83$   \\ 
        $2.5\times 10^{-4}$   & $2.2$ & $22.98$ & $2.2$ & $23.49$ & $2.2$ & $24.00$   \\ 
        $1.0\times 10^{-4}$   & $2.3$ & $24.37$ & $2.3$ & $24.74$ & $2.3$ & $25.09$   \\ 
        $1.0\times 10^{-5}$   & $2.3$ & $26.82$  & $2.3$ & $26.78$  & $2.3$ & $26.72$    
           \end{tabular}
  \caption{Critical Rayleigh numbers $\Ramc$ and their critical wavenumbers $a_{m,1}^*$ at respective $\dhatc$ values with $\epsilon_T=\{0.5, 0.7,1.0\}$ and Darcy numbers $\Da\to 0$.}
  \label{tab:critRam}
  \end{center}
\end{table}

\begin{table}
  \begin{center}
\def~{\hphantom{0}}
\begin{tabular}{l|r|r|r}
                 & \underline{$\epsilon_T=.5$}     & \underline{$\epsilon_T=.7$}     & \underline{$\epsilon_T=1.0$}     \\[3pt]
        $\sqrt{\Da}$ & \multicolumn{1}{c|}{$e_{rel}$} & \multicolumn{1}{c|}{$e_{rel}$} &  \multicolumn{1}{c}{$e_{rel}$} \\[4pt]    
$1.0\times 10^{-2}$ &        $13.00\%$&	$11.71\%$&	$10.36\%$\\
$5.0\times 10^{-3}$ & 	$9.52\%$&	$7.91\%$&	$6.55\%$\\
$2.5\times 10^{-3}$ &	$6.73\%$&	$5.33\%$&	$3.84\%$\\
$1.0\times 10^{-3}$ &	$3.89\%$&	$2.68\%$&	$1.32\%$\\
$5.0\times 10^{-4}$ &	$2.33\%$&	$1.20\%$&	$0.004\%$\\
$2.5\times 10^{-4}$ &	$1.10\%$&	$0.10\%$&	$0.96\%$\\
$1.0\times 10^{-4}$ &	$0.06\%$&	$0.91\%$&	$1.84\%$\\
           \end{tabular}
  \caption{Relative errors-- calculated with \eqref{eq:relErrorForm}-- between predicted $\dhatc$ values and values found using the linear stability analysis with $\epsilon_T=\{0.5, 0.7,1.0\}$ and Darcy numbers $\Da\to 0$.}
  \label{tab:relErrors}
  \end{center}
\end{table}

With these new critical Rayleigh numbers, we obtain the more accurate coarse-grained model for predicting the critical depth ratio:  
\begin{align}
\dhatc = \left[ \dfrac{1707.8}{27.1}\, \Da\,\epsilon_T^2 \right]^{1/4}\,. \label{eq:better}\end{align}
As we show below, the results produced with this model become increasingly accurate in the small Darcy limit, which is reasonable given that our intuition in choosing boundary conditions came from the small Darcy limit. 

\begin{figure} 
 \centering
\begin{tikzpicture}

\begin{axis}[%
width=2.0in,	
height=2.0in, 	
scale only axis,
title ={{\bf $\Da$ vs. $\dhatc$}},
ymode=log, 
    xmode=log,
    xmin=10^(-8.2),
    xmax=10^(-3.8),
    ymin=10^(-1.8),
    ymax=10^(-.4), 
    xlabel={Darcy number, $\Da$},
    ylabel= {Critical depth ratio, $\dhatc$},
    grid=major,
    extra y ticks={10^(-.5),10^(-1.5)},
    legend style={at={(1.3,.75)},
    anchor=north,legend columns=1}
]
                \addplot[firstColor, line width=1pt] 
      coordinates {
		((1.0e-02)^2, 0.199228714051074)
		((1.0e-04)^2, 0.0199228714051074)
	};
      	  \addplot[secondColor, line width=1pt] 
      coordinates {
		((1.0e-02)^2, 0.235730593482099)
		((1.0e-04)^2, 0.0235730593482099)
	};
      		 \addplot[thirdColor, line width=1pt] 
      coordinates {
		((1.0e-02)^2, 0.281751949425180)
		((1.0e-04)^2, 0.0281751949425180)
	};
		\addlegendimage{combo legend}
      \legend{ $\epsilon_{\,T}=0.5$,$\epsilon_{\,T}=0.7$,$\epsilon_{\,T}=1.0$};
      
      \addplot[black, line width=1pt] 
      coordinates {
		((1.0e-02)^2, 0.199228714051074)
		((1.0e-04)^2, 0.0199228714051074)
	};
	\label{blackLine}
	
	       \addplot[firstColor, line width=1pt] 
      coordinates {
		((1.0e-02)^2, 0.199228714051074)
		((1.0e-04)^2, 0.0199228714051074)
	};
	
	\addplot[scatter, 
      only marks, 
      scatter/use mapped color = {draw=black, fill=black},black] 
      coordinates {
		((1.0e-02)^2, .229)
		((5.0e-03)^2, .1557)
		((2.5e-03)^2, .1068) 
		((1.0e-03)^2, .06555) 
		((5.0e-04)^2, .04561)
		((2.5e-04)^2, .03185) 
		((1.0e-04)^2, .01991)
	};	
	\label{blackDot}
	
            \addplot[scatter, 
      only marks, 
      scatter/use mapped color = {draw=firstColor, fill=firstColor},firstColor] 
      coordinates {
		((1.0e-02)^2, .229)
		((5.0e-03)^2, .1557)
		((2.5e-03)^2, .1068) 
		((1.0e-03)^2, .06555) 
		((5.0e-04)^2, .04561)
		((2.5e-04)^2, .03185) 
		((1.0e-04)^2, .01991)
	};		  
	\addplot[scatter, 
      only marks, 
      scatter/use mapped color = {draw=secondColor, fill=secondColor}, secondColor] 
      coordinates {
		((1.0e-02)^2, .267)
		((5.0e-03)^2, .181)
		((2.5e-03)^2, .1245) 
		((1.0e-03)^2, .0766) 
		((5.0e-04)^2, .05335)
		((2.5e-04)^2, .03731) 
		((1.0e-04)^2, .02336)
	};
      \addplot[scatter, 
      only marks, 
      scatter/use mapped color = {draw=thirdColor, fill=thirdColor}, thirdColor] 
      coordinates {
		((1.0e-02)^2, .3143)
		((5.0e-03)^2, .2132)
		((2.5e-03)^2, .1465) 
		((1.0e-03)^2, .09029) 
		((5.0e-04)^2, .063004)
		((2.5e-04)^2, .044125) 
		((1.0e-04)^2, .027666)
	};
	 \addplot[black, line width=1pt] 
      coordinates {
		(3.16228e-07, 0.0316227766)
		(0.0000316228, 0.0316227766)
		(0.0000316228, 0.1)
		(3.16228e-07, 0.0316227766)
	};
	    \addplot [black, line width=1pt] coordinates {(3.16228e-07, 0.0316227766) (0.0000316228, 0.0316227766)}
             node[below, pos = .5] {4};
             \addplot [black, line width=1pt] coordinates {(0.0000316228, 0.0316227766) (0.0000316228, 0.1)}
             node[right, pos = .5] {1};
\end{axis}

\node[draw,fill=none,right] 
	at(5.53,1.6) {\begin{tabular}{cl} \ref{blackLine} & Marginal Stability \\ \ref{blackDot}& Theory \end{tabular}};

\end{tikzpicture}%
\caption{Critical depth ratios for various $\epsilon_T$ (ratio of thermal diffusivities) values, $\epsilon_T=0.5, 0.7, 1.0$. The solid lines represent the predicted $\hat{d}^*$ values from our theory \eqref{eq:better}, and the circles are the $\hat{d}^*$ values calculated from the marginal stability curves determined by \cite{mccurdy2019convection}. \label{fig:dhatc}}
\end{figure}
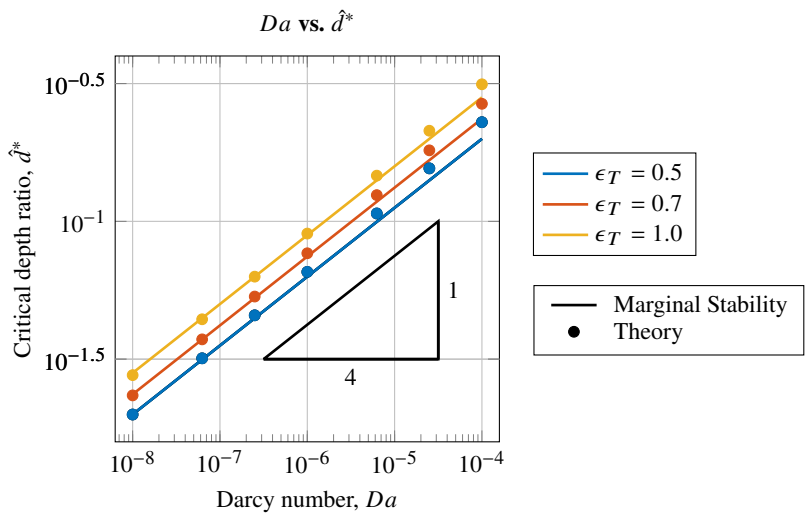

Figure~\ref{fig:dhatc} shows data for three different $\epsilon_T$ values while varying $\Da$, since our formula is a function of these two variables. As $\Da\to0,$ the critical depth ratio $\dhatc$ goes to zero as well.
We plot the critical depth ratios found with two methods-- the first predicted from the heuristic theory, compared to the second obtained from the marginal stability results from \cite{mccurdy2019convection}. 
To show that our model predicts $\dhatc$ going to zero at the same rate as the actual values found using the linear stability analysis $\dhatc_{LSA}$, we plot the data $\Da$ vs. $\dhatc$ with a log-log plot along with a reference triangle to illustrate the slope of $1/4.$

To quantify the error of our predicted $\dhatc$ values, we define the relative error between the predicted $\dhatc$ values from \eqref{eq:better}, $\dhatc_{Thry}$, and values found using the linear stability analysis, $\dhatc_{LSA}$, with \begin{align}
e_{rel} = \frac{\left|\dhatc_{LSA}-\dhatc_{Thry} \right|}{\dhatc_{LSA}} \label{eq:relErrorForm}\,.
\end{align} 
The relative errors are noted in table \ref{tab:relErrors}. For $\Da=1.0\times10^{-4}$ values, the model has about 10\% relative error, while the relative error drops to less than 2\% when $\Da=1.0\times10^{-8}.$ 
The worst relative error with our new formula outperforms the best relative error of the formula presented in \cite{mccurdy2019convection}.

This coarse-grained model allows us to accurately predict the depth ratio that triggers a shift between deep and shallow convection. 
In applications with technology, being able to harness convection in this manner could have considerable impact with the design of heat sinks. 
By selecting physical properties of the porous medium, a suitable depth ratio could be chosen to obtain the desired flow configuration to prevent overheating and enhance circulation. 
Rather than conduct costly experiments to determine a depth ratio that yields the intended results, our formula provides a reference for an appropriate range of $\dhat$ values needed for deep or shallow convection.

This bifurcation from shallow to deep convection is highlighted in several works, using linear stability or numerical simulations to show the deep versus shallow convection profiles. Finding the critical depth ratio with either of these methods-- linear stability and numerical simulations-- can be computationally-expensive and time-consuming. Our model allows us to accurately predict a range of depth ratios where this shift in convection occurs, each of which are in good agreement with the critical depth ratios found by previous researchers. For example, with the same governing equations we consider, \citet[][]{chen1988} showed via linear stability that a transition occurred between depth ratios of $\dhat = 0.12$ and $\dhat=0.13$ using the fixed parameters $\sqrt{Da}=0.003,\,\epsilon_T=0.7$. Substituting these values into \ref{eq:better} allows our model to predict that a transition occurs at the depth ratio of 
\begin{align*}
\dhatc = \left[ \dfrac{1707.8}{27.1}\, \left(0.003\right)^2\,\left(0.7\right)^2 \right]^{1/4}\approx 0.1291\,,\end{align*} agreeing well with the work conducted by Chen \& Chen. The recent work by \citet[][]{han2020dynamic} presented an example where the convection profiles shifted from shallow to deep convection as the depth ratio was altered from $\dhat=0.16$ to $\dhat=0.17$, found via numerics and their center manifold theory (see figure 5.2 of \cite{han2020dynamic}). With the values of $\Da=25\times 10^{-6}$ and $\epsilon_T=0.7$ from this example in their work, we predict a critical depth ratio of $\dhatc \approx \,0.1667\,,$ which falls in the range Han {\em et. al} determined. The formula we developed greatly reduces the computational time needed to determine where the shallow to deep transition occurs. Using our coarse-grained model bypasses the necessity of searching through large parameter spaces while solving stability problems or conducting numerical simulations, as it quickly narrows the parameter regime where the transition occurs, especially for small Darcy numbers. In Appendix \ref{appA}, we present data showing good agreement between true critical depth ratios and those predicted from our model. Additionally, we show that our formula \eqref{eq:better} can be used to predict how altering the ratio of thermal diffusivities $\epsilon_T$ or the Darcy number $\Da$ can trigger a shift in convection.

Although each of the papers noted above use 2D simulations or experiments in their work, our model can also be used in 3D settings as well. Since our theory comes from the stability of the uncoupled systems under the assumption of infinite horizontal plates, both the 2D and 3D problems reduce to one spatial dimension-- the vertical direction. As such, our model can be applied to both 2D and 3D systems, albeit with some error since the assumption of infinite horizontal plates cannot be satisfied in physical settings or numerical simulations.

The main limitation of our coarse-grained model is that the results are confined to the {\em onset} of convection, where Raleigh numbers are close to their critical counterparts. 
To develop a more comprehensive understanding of this phenomenon over a more broad parameter regime, we turn to numerical simulations of the full, nonlinear system.

\section{Numerical Scheme}
\label{sec:numerics}

The stability analyses and coarse-grained model can predict flow configurations near the onset of convection, but numerical simulations are needed to examine dynamics outside of this regime.
In this section, we present a numerical scheme to simulate convection in the superposed fluid-porous medium system using a FEM. 
The main advantage to this scheme is that, by lagging the nonlinear terms and the terms associated with interface conditions, we produce a set of linear, sequentially decoupled equations.
A similar scheme was studied by \citet[][]{chen2020uniquely} with the Cahn-Hilliard-Navier-Stokes-Darcy-Boussinesq system-- or, the same system we study with the inclusion of a phase-field function using the Cahn-Hilliard equations. 
Using a numerical method related to that of \citet[][]{chen2020uniquely} is beneficial since the method is unconditionally long-time stable. 
 We outline the scheme below, and note the time-lagged terms as they are shown. 
 Next, we detail how the variational forms are obtained, using the following notation for vector-valued functions $\vecf$ and $\vecg$ and matrix-valued functions $\matA$ and $\matB$:
    \begin{align*}
    \left(\vecf, \vecg \right)_j = \int_{\Omega_j}\vecf\bcdot \vecg\,d\Omega_j\, , \hspace{.15in}\langle\matA, \matB \rangle_j = \int_{\Omega_j}\matA\,\colon\matB\,d\Omega_j\, , \hspace{.15in} \|\vecf\,\|_j^2=(\vecf,\vecf\,)_j\, , \hspace{.15in} |\vecf\,|^2 = \vecf\bcdot \vecf\, ,
    \end{align*}
for domains $j \in \{f,m\}$ where the $f$ and $m$ subscripts correspond to the fluid and porous medium regions, respectively. 

We introduce the following FE spaces:
\begin{itemize}
    \item $W_f = \{ \wuf \in \left[H^1\left(\Omega_f \right)\right]^2: \wuf \bcdot \boldsymbol{n} =0 \textrm{ at top $+$ periodic on left and right} \}\,$,\\[-5pt]
    \item $Q_f = \{ q_f \in L^2\left(\Omega_f \right): \int_{\Omega_f} q_f\,d\boldsymbol{x}=0 \}=L^2_0\left( \Omega_f\right)\,$,\\[-5pt]
    \item $Q_m = \{ q_m \in L^2\left(\Omega_m \right): \int_{\Omega_m} q_m\,d\boldsymbol{x}=0 \}=L^2_0\left( \Omega_m\right)\,$,\\[-5pt]
    \item $\Psi = \{ \psi \in H^1\left(\Omega \right): \psi =0 \textrm{ at top and bottom $+$ periodic on left and right} \}\,$.\\[-5pt]
\end{itemize} 
The space $\Psi$ spans the entire domain and is reserved for the temperature field over $\Omega$, since the advection-diffusion equation (ADE) for heat can be written as one equation over the entirety of the domain. 
Instead of solving two problems for $\theta_f$ and $\theta_m,$ we solve only one for 
\begin{equation}
  \theta = \left\{
    \begin{array}{ll}
      \theta_f &\textrm{for } \boldsymbol{x}\in\Omega_f\,, \\[2pt]
      \theta_m & \textrm{for }\boldsymbol{x}\in\Omega_m\,.
    \end{array} \right.
\end{equation}
With superscripts denoting the time iteration, we present the variational problems corresponding to the system  \eqref{eq:NavStokesSysNondim}--\eqref{eq:BCnondim}.

First, given $\left(\vufp^{(n)},\vufp^{(n-1)}, \theta_m^{(n)}\right)$, we find the perturbed pressure in the porous medium $\pi_m^{(n+1)}\in Q_m$ with 
 \begin{align}
&\Da\left(\bnabla \pi_m^{(n+1)}, \bnabla q_m\right)_m -\Ram\, \left(\theta_m^{(n)}\boldsymbol{k}, \bnabla q_m\right)_m \nonumber  \\[5pt]
&\quad +\int_{\Gamma_i}\left[ \dfrac{\Da}{\Prm\, \chi }\,\dfrac{\vufp^{(n)}-\vufp^{(n-1)}}{\Delta t} + \vufp^{(n)} \right] \bcdot \boldsymbol{n}\, q_m \,d\Gamma_i=0\quad \forall q_m\in Q_m\,.\label{eqn:variDarcy}
\end{align} 
 To begin decoupling the problems, we use the previous $\vufp^{(n)}$ values in the integral along the interface. 
 Our treatment of the interface term, along with the method in which we solve for the Darcy velocity, differ from the method used in \citet[][]{chen2020uniquely}.
 We still observe experimentally that our method is stable, with the time-step restriction of $\Delta t \sim \mathcal{O}(\Da)$ or smaller. 
Having solved for $\pi_m^{(n+1)}$, we can use the previous Darcy velocity $\vump^{(n)}$ to find the updated velocity $\vump^{(n+1)}$. 
The Darcy equation is 
\begin{align*}
 \dfrac{\Da}{\Prm\, \chi}\dfrac{\vump^{(n+1)} - \vump^{(n)} }{\Delta t} + \vump^{(n+1)} &=-\Da\bnabla \pi_m^{(n+1)} +\Ram\,\theta_m^{(n)}\,\boldsymbol{k} \, ,
\end{align*}
which we can solve for $\vump^{(n+1)}$ with
\begin{align}
\vump^{(n+1)} &=\left[-\Da\bnabla \pi_m^{(n+1)} +\Ram\,\theta_m^{(n)}\,\boldsymbol{k} + \dfrac{\Da}{\Prm\, \chi\,\Delta t} \vump^{(n)}\right]\left(\dfrac{\Prm\, \chi\,\Delta t}{\Da +\Prm\, \chi\,\Delta t} \right) \, . \label{eqn:updateDarcyVel}
\end{align} 

Next, using $\left(\vufp^{(n)},\,\pi_m^{(n+1)},\,\theta_f^{(n)}\right)$--where the previously-found $\pi_m^{(n+1)}$ term is used in an integral along the interface-- we find $\left(\vufp^{(n+1)},\,\pi_f^{(n+1)}\right)\in W_f\times Q_f$ by solving
\begin{align}
 &\left(\frac{1}{\Prm}\,\dfrac{\vufp^{(n+1)}-\vufp^{(n)}}{\Delta t},\wuf \right)_f+ \mathsfbi{B}_f\left(\vufp^{(n)},\vufp^{(n+1)},\wuf \right)+2 \left<\mathsfbi{D}\left(\vufp^{(n+1)} \right),\mathsfbi{D}\left(\wuf \right) \right>_f \nonumber \\[5pt]
 &\quad - \left(\pi_f^{(n+1)},\bnabla \bcdot \wuf \right)_f + \left(\bnabla \bcdot \vufp^{(n+1)}, q_f \right)_f +\dfrac{\Ram}{\Da} \left( \theta_f^{(n)}, \wuf\bcdot \boldsymbol{k} \right)_f +  \int_{\Gamma_i}\pi_m^{(n+1)} \left(\wuf \bcdot \boldsymbol{n} \right) \,d\Gamma_i  \nonumber \\[5pt]
 &  \quad+  \int_{\Gamma_i}\dfrac{\alBJSJ}{\sqrt{\Da}}  \left(\vufp^{(n+1)} \bcdot \boldsymbol{\tau} \right)\left(\wuf \bcdot \boldsymbol{\tau} \right) \,d\Gamma_i= 0 \quad \forall \wuf\in W_f,\,q_f\in Q_f\,, \label{eqn:variNS} 
\end{align} where the trilinear term $\mathsfbi{B}_f$ is defined with 
\begin{align*}
2\,\mathsfbi{B}_f\left(\vu,\vup, \wu \right) =  \int_{\Omega_f}\left(\vu\bcdot \bnabla \vup \right)\wu - \left(\vu\bcdot \bnabla \wu \right)\vup\,d\Omega_f + \int_{\Gamma_i}\left(\vu\bcdot \wu \right)\left(\vup \bcdot \boldsymbol{n} \right) - \left(\vu\bcdot \vup \right)\left(\wu \bcdot \boldsymbol{n} \right)\,d\Gamma_i\,.
\end{align*} The first integral of the trilinear term $\mathsfbi{B}_f$ was first used by Temam in the late 1960s \citep[][]{temam1968methode,temam1969approximationOne,temam1969approximationTwo} and has remained a critical tool in approximating solutions to the Navier-Stokes equations since then. The second integral is included to deal with the non-homogeneous boundary value in our application. The skew-symmetric form of the trilinear term $\mathsfbi{B}\left(\vufp^{(n)},\vufp^{(n+1)},\wuf \right)_f$ allows for partial time-lagging of the nonlinear term of Navier-Stokes, which linearizes the problem while still conserving the energy of the system. Treating the nonlinear term in this manner has been used in a number of recent works \citep[][]{chen2020uniquely,chen2021conservative}, albeit for an additional reason as well. With the coupled Navier-Stokes-Darcy equations, use of this trilinear form allows for cancellation of the $\frac{1}{2}|\vufp|^2$ term from the Lions interface condition \eqref{eq:InterfCondNondim.5} with integration by parts on the $(\vup\bcdot\bnabla)\vup$ term of Navier-Stokes. 


With the velocity of both sub-domains known, we write them as a single, updated velocity field:
\begin{equation}
  \vup^{(n+1)} = \left\{
    \begin{array}{ll}
      \vufp^{(n+1)} &\textrm{for } \boldsymbol{x}\in\Omega_f\,, \\[2pt]
     \vump^{(n+1)} & \textrm{for }\boldsymbol{x}\in\Omega_m\,.
    \end{array} \right.
\end{equation} With $\left(\vup^{(n+1)},  \theta^{(n)}\right)$, we obtain $\theta^{(n+1)}\in \Psi$ by solving
\begin{align}
  & \left(\delta_1\,\dfrac{\theta^{(n+1)}-\theta^{(n)}}{\Delta t}, \psi \right)_\Omega + 
   \Prm\, \mathsfbi{B}\left(\vup^{(n+1)},\theta^{(n+1)} \, , \psi \right) \nonumber \\
   &\quad + \left(\delta_2 \bnabla \theta^{(n+1)} \,,\bnabla \psi \right)_\Omega - \left(\delta_3  \vup^{(n+1)}\bcdot \boldsymbol{k} \,, \psi \right)_\Omega  =0\,\quad \forall \psi\in\Psi\label{eqn:variHeat}
\end{align}  
where the coefficients $\delta_i$ are defined by 
\begin{align}
    \delta_1=\begin{cases}
    1 &\textrm{for } \boldsymbol{x}\in\Omega_f\,, \\[2pt]
    \varrho & \textrm{for }\boldsymbol{x}\in\Omega_m\,,
    \end{cases} \quad 
    \delta_2=\begin{cases}
    \epsilon_T &\textrm{for } \boldsymbol{x}\in\Omega_f\,, \\[2pt]
     1 &\textrm{for } \boldsymbol{x}\in\Omega_m\,,
    \end{cases} \quad 
    \delta_3=\begin{cases}
    1/\epsilon_T &\textrm{for } \boldsymbol{x}\in\Omega_f\,, \\[2pt]
    1 & \textrm{for }\boldsymbol{x}\in\Omega_m\,,
    \end{cases}\label{eqn:consts}
    \end{align} 
    and the trilinear term $\mathsfbi{B}$ is defined with 
\begin{align*}
2\,\mathsfbi{B}\left(\vu, v, w \right) =  \int_{\Omega}\vu\bcdot\left(w\bnabla v\right) - \vu\bcdot\left(v\bnabla w\right)\,d\Omega\,.
\end{align*} 
    These coefficients are discontinuous over the interface since they are related to physical properties of each region, and are chosen to enforce the interface conditions-- continuity of the temperature and heat flux. 
    With the updated velocity field over the entire domain, we are also able to solve for the streamlines $\phi$:
\begin{align}
   \left( \bnabla\phi^{(n+1)}, \bnabla \varphi\right)_\Omega - \left(\bnabla \times \vup^{(n+1)}, \varphi \right)_\Omega  = 0\, \quad  \forall \varphi \in \Phi\,,\label{eqn:variStream}
\end{align} 
where $\Phi=\{\varphi \in H^1\left(\Omega \right): \varphi=0 \textrm{ on top and bottom $+$ periodic on left and right}\}$.

Each simulation begins by perturbing the conductive steady-state. That is, we apply a seeded, random perturbation (on the order of $10^{-6}$) to a stationary fluid and a piecewise linear temperature profile. The simulations begin with:
 \begin{align}
   \vufp^{(0)} &= \vump^{(0)} = \boldsymbol{0} + \boldsymbol{\varepsilon}_1(\boldsymbol{x})\, , \nonumber \\
   \theta_f^{(0)}&=\theta_m^{(0)} = 0 + \varepsilon_2(\boldsymbol{x})\,,
        \end{align}
where $\boldsymbol{\varepsilon}_1(\boldsymbol{x})$ and $\varepsilon_2(\boldsymbol{x})$ are small, seeded, random perturbations. These perturbations-- the components of the vector-valued function $\boldsymbol{\varepsilon}_1(\boldsymbol{x})$ and the scalar $\varepsilon_2(\boldsymbol{x})$-- take the form \begin{align*}
  A\sum_{k,\ell=-N}^{N}\frac{1}{\sqrt{\pi}}e^{-\frac{1}{2}\left(k^2+\ell^2\right)}\sin\left(\frac{k\pi\,x}{L_x} + \alpha_{k,\ell}\right) \sin\left(\frac{\ell\pi\,(y-\hat{d}\,)}{1+\hat{d}} + \beta_{k,\ell}\right)\,,
    \end{align*} where $\alpha_{k,\ell},\,\beta_{k,\ell}$ are randomized phases from a uniform distribution on $[0,2\pi)$ and $A$ sets the overall magnitude of the perturbation.

For numerical stability, 
the time-step is heavily restricted by the requirement $\Delta t \sim \mathcal{O}(\Da)$. 
Our method is first-order in time and second-order in space. 
The time-integrator could be swapped out for a high-order method. 
However, in practice, we find that the spatial error dominates errors from time discretization, essentially rendering higher-order time integrators unnecessary. 
Each simulation conducted in this work has a length scale in $x$ of $2.1$ units, which, through experimentation, we determined to be the smallest domain able to support a pair of counter-rotating, deep convection cells.
The length scale in $y$ has height $1+\dhat,$ with $1$ unit for the porous height and $\dhat$ for the fluid region. 
For the spatial discretisation, we use a uniform grid with $32 \times 32$ triangular elements over a unit area in the nondimensional domain. Future work is being conducted on using a nonuniform mesh for this system to investigate the associated errors with this discretisation.
All of the numerical simulations are run using FreeFem++ \citep[][]{freefemCite}.

\section{Results and Discussion}
\label{sec:results}

\begin{figure} 
 \centering
\input{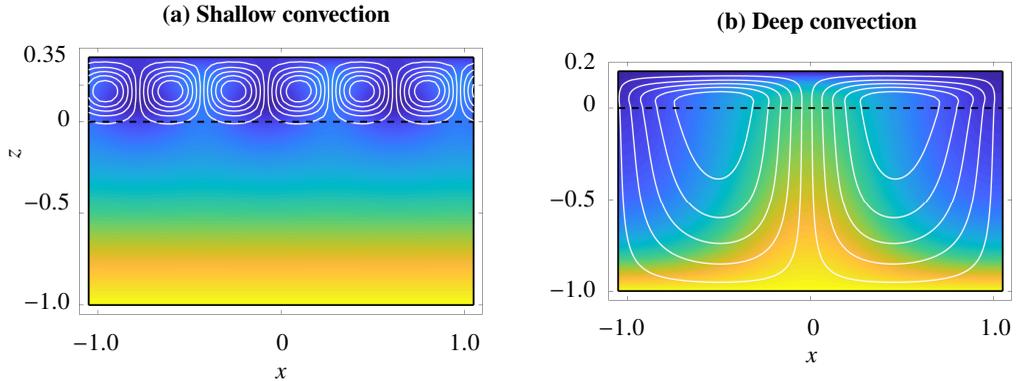}\\[\baselineskip \vspace{-.20in}]
\caption{Shallow and deep convection with temperature (color) and streamlines (contour) at their respective steady-states, with (a) $\Ram=10,\,\dhat=.35$ at $t=3.0$ and (b) $\Ram=30,\,\dhat=.20$ at $t=3.0$, respectively. Fixed parameters: $\Da=1.0\times10^{-4},\,\Prm =.7,\,\epsilon_T=.7,\,\alpha=1.0,\, \Delta t = 2.5\times10^{-4}.$\label{fig:deep_shallow_cases}}
\end{figure}

From stability analyses and our theory, there are two types of convection we expect-- deep and shallow convection. 
In figure~\ref{fig:deep_shallow_cases}, these two flow configurations are shown at their steady-states with streamlines shown as countours and the temperature profiles in color. 
The system with deep convection exhibits cells which extend throughout the entirety of the domain, evidenced by the wave-like temperature profile and streamlines circulating throughout the whole domain.
In contrast, with shallow convection, almost all of the velocity and temperature deviations from the conductive state are confined to the fluid region alone, leaving the porous medium largely unchanged with the exception of a small region immediately below the interface. 
For these two cases, the systems begin with random initial data and go directly to their preferred convection states.

With the case shown in figure~\ref{fig:deep_shallow_cases}a, the fluid evolves from random, disorganized data to a steady-state with shallow convection.
Conversely, for the simulation in figure~\ref{fig:deep_shallow_cases}b, the system goes from the perturbed conductive state to a stable steady-state of deep convection.
That is, in these two respective cases, the shallow convection case's cells originate and stay in the fluid region, while cells from the deep convection case occupy the whole domain for the duration of the simulations.
This phenomenon is expected; with these two types of convection, instabilities from one group of wavenumbers outperform the other by a significant margin. 
For example, with deep convection, the growth from small wavenumbers of the perturbation dominates any contribution from the larger wavenumbers, resulting in large cells. 
Alternatively, with shallow convection, the growth from small wavenumbers pales in comparison to that of the larger wavenumbers.

\begin{figure}
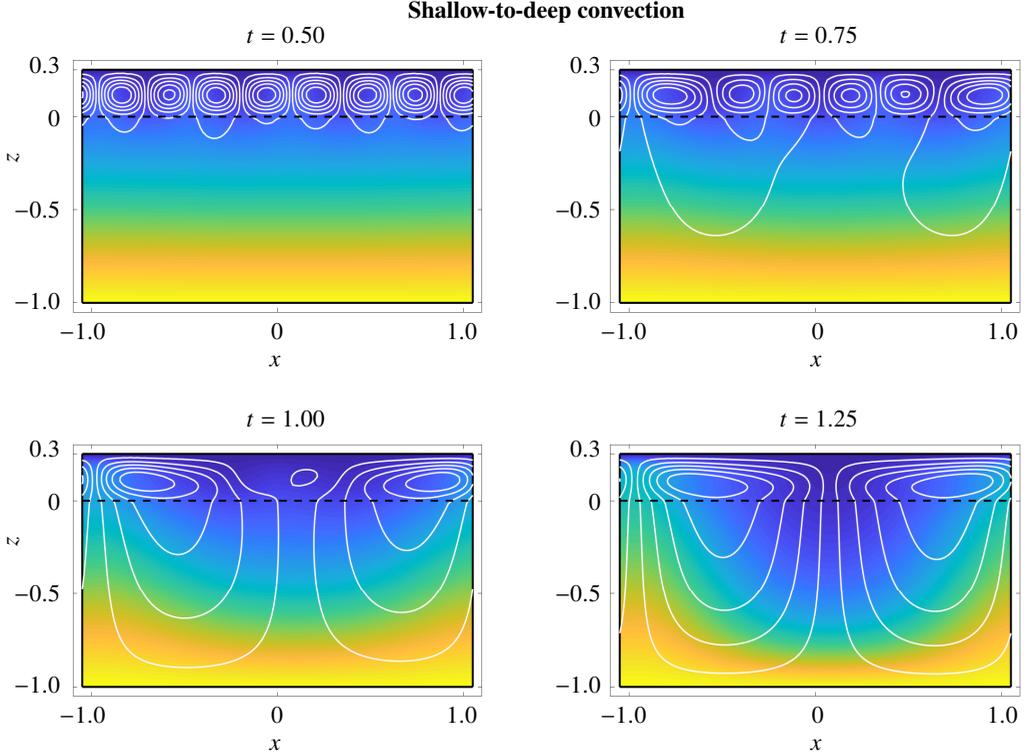
 
\centering
\begin{subfigure}{\linewidth}
   \centering
  \input{tikz_files/fig5ab}
\end{subfigure}
\\[\baselineskip \vspace{-.15in}]
\begin{subfigure}{\linewidth}
 \centering
\input{tikz_files/fig5cd}
\end{subfigure}
\\[\baselineskip \vspace{-.35in}]
\centering
\caption{Shallow-to-deep convection with $\Ram=20,\,\dhat=.30$ with temperature (color) and streamlines (contour). Fixed parameters: $\Da=1.0\times10^{-4},\,\Prm =.7,\,\epsilon_T=.7,\,\alpha=1.0,\, \Delta t = 2.5\times10^{-4}.$ \label{fig:transition_case}}
\end{figure}

There is another form of convection that the stability analyses cannot predict though: shallow-to-deep convection, as shown in figure~\ref{fig:transition_case}. 
This kind of convection takes place in a parameter regime where {\em both} the small and large wavenumbers are unstable, and the interaction between their instabilities gives rise to this hybrid category of convection. 
In these situations, the system goes from random initial data to a metastable shallow convection state, followed by the collapse and consolidation of cells in the fluid region. 
The joined cells gain enough strength to penetrate into the porous medium, forming larger-scale cells which occupy the fluid and porous regions alike, and the system finally steadies out at the preferred stable configuration of deep convection. 
For this choice of parameters, this route-- from random initial data to a metastable shallow state followed by a dynamic shift to deep convection-- is {\em the} route taken from perturbing the unstable conductive state. While we observe shallow-to-deep convection, we never observe a deep-to-shallow dynamic shift; this leads us to speculate that when both small and large wavenumbers are unstable, deep convection is the preferred steady-state of this system.

Although shallow-to-deep convection and deep convection ultimately both achieve a steady-state with cells that extend throughout the domain, they are distinct flow progressions. Their most notable differences concern the routes taken to their steady-state and the efficiency of heat transfer; shallow-to-deep convection has metastable shallow cells before forming deep convection cells for its steady-state, and these flow profiles can exhibit higher Nusselt values compared to their counterparts with deep convection.

To measure convection, we define two quantities. 
The first is a mathematical energy that allows us to find the $\Ram$ value where $\frac{dE}{dt}= 0$, distinguishing between the regions of stability and instability-- with $\frac{dE}{dt}< 0$ and $\frac{dE}{dt}> 0$, respectively-- like done in \citet[][]{mccurdy2019convection}. 
This uses the volume-averaged norms of the perturbed velocity and temperature profiles, measuring how these profiles compare to their conductive steady-states:
 \begin{align} 
2\,E(t) = \frac{1}{\Prm}\frac{1}{V} \|\vup \|_{\Omega}^2 +\frac{1}{V} \|\theta\|_{\Omega}^2\,,
\end{align} with $V$ as the volume. 
Qualitatively, the onset of convection is noted by a jump in the energy profile. 
Second, we use the physically-motivated Nusselt number $Nu(t)$ as the ratio of vertical convective and conductive fluxes with
    \begin{align}\label{NusseltInfo}
       Nu(t) = \frac{ J_{cnv } + J_{cnd}}{J_{cnd}}  \quad \textrm{ with  }   J_{cnv} = \left(\vu\cdot \boldsymbol{k},\,\left(\theta + \bar{T}\right)\right)_{\Omega} \textrm{   and  } J_{cnd}=-\left(\delta_2\,\nabla \left(\theta + \bar{T}\right) \cdot \boldsymbol{k},1\right)_{\Omega}\,.
    \end{align} where $\bar{T}$ is the conductive temperature and $\delta_2$ is defined in \eqref{eqn:consts}, which allows us to write the conductive fluxes of each region as a single term.

\begin{figure} 
\centering
\input{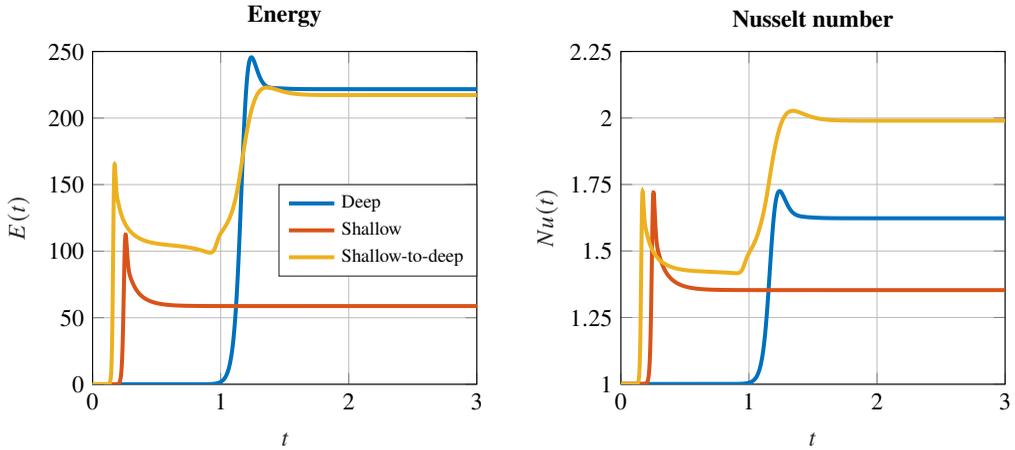}
 \caption{Energy and Nusselt profiles for the three flow configurations shown in figures \ref{fig:deep_shallow_cases} and \ref{fig:transition_case}. Fixed parameters: $\Da=1.0\times10^{-4},\,\Prm =.7,\,\epsilon_T=.7,\,\alpha=1.0,\, \Delta t = 2.5\times10^{-4}.$ \label{fig:energyPlot}}
\end{figure}

In figure~\ref{fig:energyPlot}, we plot the energy and Nusselt profiles for the three different qualitative states presented in figures \ref{fig:deep_shallow_cases} and \ref{fig:transition_case}. 
For each curve, the initial jump indicates the time of noticeable formation of convection cells in their respective cases with their steady-states achieved as the profiles level out. 
The deep and shallow cases (shown in red and blue, respectively) show the formation of cells, immediately followed by a steady-state. 
The phenomenon of shallow convection cells forming faster than deep cells is expected. 
Shallow cells have nothing obstructing their formation, while deep convection is hindered by the porous medium. 
As a result, it takes longer for the fluid to gain enough velocity in the porous region for noticeable deep convection cells to form. 
More precisely, the characteristic timescale for the porous-media flow given in \eqref{ScalesEqn} is ${d_m^2}/{\lambda_m}$, which has been normalized to unity in our analysis. The flow timescale for the fluid region, meanwhile, is ${d_f^2}/{\lambda_f}$, which is shorter by a factor of $\dhat^2/\epsilon_T$.

These timescales can be used to better understand the shallow-to-deep convection case, which is shown by the yellow curves in figure~\ref{fig:energyPlot}. 
The first jump in $E$ and $Nu$ occurs as the convection cells form in the free-flow zone, followed sometime later by a secondary jump whereby the cells coalesce into the larger, deep convection cells and push into the porous medium. 
As noted above, the timescale for porous-media flow has been normalized to unity and the timescale for free-fluid flow is shorter by a factor of $\dhat^2/\epsilon_T$ or about 0.09 in the case shown. 
These values are consistent with the quick initial onset free-flow convection, followed by a slower shift to deep convection.

The jumps in $Nu$ with the shallow-to-deep case demonstrate that these configurations have the potential to transfer heat more efficiently than their counterparts with deep or shallow convection. 
Further investigations into this hybrid flow pattern could determine the parameter regimes which produce shallow-to-deep convection to maximize heat flow in various applications, like heat sink technology. 
For example, in comparing the deep convection and shallow-to-deep convection cases with the $E(t)$ and $Nu(t)$ profiles, we see both cases eventually steady-out to approximately the same energy; that is, they have comparable deviations from their conductive states. 
However, the case with shallow-to-deep convection has a larger Nusselt number than its deep-convection counterpart, indicating a greater efficiency in transferring heat. We speculate that shallow-to-deep convection is more efficient in its heat transfer than deep convection due to the route each takes to its steady state. For shallow-to-deep convection, as cells form and gain speed in the free-flow region, there is a spike in the the $Nu$ profile. As the shallow cells consolidate and penetrate into the medium, a second spike in the Nusselt value occurs.

The parameters for the simulations with deep and shallow convection are chosen so that a single group of wavenumbers was the contributing to the perturbations. 
These calculations come from looking at the marginal stability curves that identify the $\Ram$ values for the onset of convection at a given depth ratio. 
For larger Rayleigh numbers, where the small and large wavenumbers both contribute the perturbation growth, the stability analyses cannot predict the preferred flow configuration. 
To provide a more holistic picture of flow behaviour, we use marginal stability analyses and numerical simulations to identify regions in a $\dhat-\Ram$ phase space where each configuration occurs.

\begin{figure}
 \centering
\input{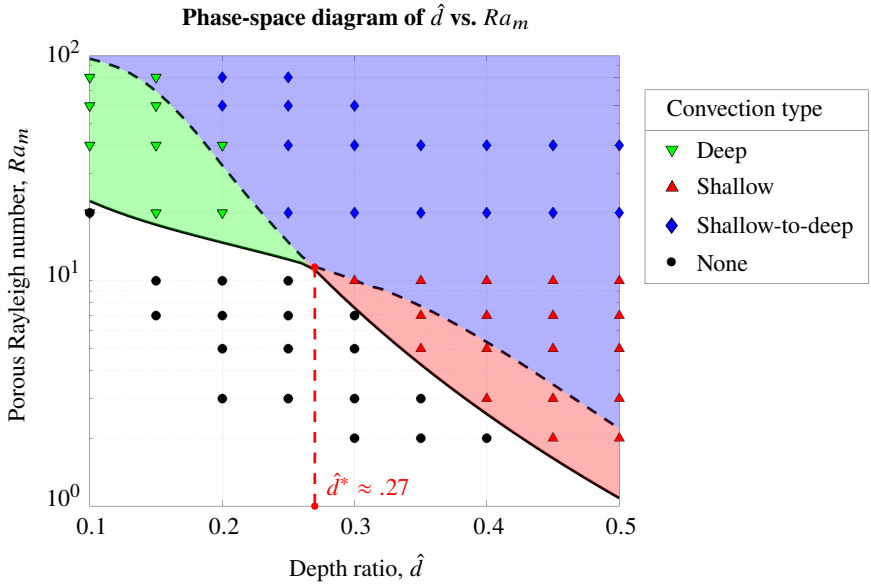}\\[\baselineskip \vspace{-0.12in}]
 \caption{Phase space with different types of convection noted. Solid black lines note the $\Ram$ values for marginal stability, and dashed lines mark the $\Ram$ values where both large and small wavenumbers become unstable. Fixed parameters: $\Da=1.0\times10^{-4},\,\Prm =.7,\,\epsilon_T=.7,\,\alpha=1.0,\, \Delta t = 2.5\times10^{-4}.$\label{fig:phase_space}}
\end{figure}

Marginal stability analyses are able to predict the $\Ram$ values for the onset of convection at a given depth ratio (shown with the solid black lines of figure~\ref{fig:phase_space}). 
There is a small region above the critical Rayleigh numbers where a single group of wavenumbers is unstable, resulting in either deep or shallow convection, and the upper bound is noted in the phase-space diagrams with the dashed lines. 
For example, with $\dhat = .2$ and $\Ram=25$, only the small wavenumbers are unstable, resulting in deep convection. 
However, once the Rayleigh number is increased passed the dashed line, say to $\Ram=50$ or $\Ram=80$, the larger wavenumbers are unstable as well.
In regions where both groups of wavenumbers unstable, linear stability analysis and our coarse-grained model are not able to accurately predict convection patterns.
In practice, we find the deep or shallow convection regimes extend above the predicted regions from the stability analyses, as evidenced with the green and red triangles extending into the blue region where both groups of wavenumbers are unstable. The distance in which the red triangles persist into the blue region tells us that for these Rayleigh numbers, the small wavenumbers (which produce the deep convection cells) are only slightly unstable, and not enough so to overpower the instabilities coming from the large wavenumbers which produce the shallow cells. 
For Rayleigh numbers well-above these regions in the phase space, where larger and smaller wavenumbers can interact and our analytical tools are ineffective, shallow-to-deep convection takes place.

The phase space shown in figure~\ref{fig:phase_space} ties our theory, the stability analyses, and numerical simulations together nicely. 
Marginal stability predicts the Rayleigh numbers needed for the onset of convection, and the wavenumbers associated with the unstable modes ultimately dictate the flow configuration for Rayleigh numbers near their corresponding critical value. 
This is represented in the phase diagram by the region between the solid and dashed lines where the marginal stability curves can effectively determine the behaviour of convection. 
The shift in the unstable wavenumbers occurs where the solid and dashed lines come together, noting the shift in convection, that can accurately be determined by our theory. 
The numerical simulations agree with results from the stability analyses and theory for Rayleigh numbers around the onset of convection. 
Further, the numerics allow for a more comprehensive understanding of superposed fluid-porous convection with higher Rayleigh numbers.

\section{Conclusions}
Based on the framework developed in \citet[][]{mccurdy2019convection}, we proposed a coarse-grained model to predict the critical depth ratio needed for the transition from deep to shallow convection, and we observed the formula is increasingly accurate in the limit of $\Da\rightarrow 0$. 
The previous stability analyses and the new transition theory, when viewed in tandem with these novel numerical simulations, provide a more complete view into the phenomenon of convection in superposed fluid-porous media systems.

The coarse-grained model presented was formed under the assumption that heuristically, the shift between deep convection and shallow convection takes place when the Rayleigh numbers of the fluid and porous regions are equivalent in some sense. 
In the physically relevant small Darcy number regime, we deduced appropriate boundary conditions for the uncoupled regions to determine their critical Rayleigh numbers, and as a result, we saw the formula became more accurate in the small Darcy limit of $\Da\rightarrow 0.$

To explore the extensive parameter regime outside the limitations of our model, we outlined a numerical scheme to simulate the full nonlinear system. 
The main contributions were with the decoupling and time-lagging of nonlinear terms; these adjustments to the system allow for linear, decoupled, sequential solvers to be used in approximating solutions.

With our simulations, we verified results from our coarse-grained model and previous stability analyses, namely with deep and shallow convection. 
However, we additionally observed a type of convection not able to be predicted by marginal stability, shallow-to-deep convection. 
This kind of convection presents an exciting new direction of work, as there is potential to investigate the dynamic reorganization of the flow. 
Additionally, the numerical simulations provide opportunities to explore less structured domains, perhaps with more complex interfaces \citep[][]{allen1984, han2016decoupled}, boundaries that evolve in time \citep[][]{zhang2000periodic, MooreCPAM2017, Quaife2018, chiu2020viscous, mac2022morphological}, or with some permeable membrane developing between the regions \citep[][]{eastham2020multiphase, sorribes2019biomechanical}.

\backsection[Acknowledgements]{We would like to thank the reviewers for carefully critiquing the manuscript. With their suggestions, we were able to better convey our arguments and present our results in a clearer manner.}

\backsection[Funding]{This work was supported by the National Science Foundation (N.M., grant DMS-2012560) and the National Natural Science Foundation of China (X. W., grants 12271237 and 11871159).}

\backsection[Declaration of interests]{The authors report no conflict of interest.}

\backsection[Data availability statement]{The data that support the findings of this study are openly available at http://doi.org/10.5281/zenodo.7320220.}

\backsection[Author ORCIDs]{M. McCurdy, https://orcid.org/0000-0002-7343-1659; N. Moore, https://orcid.org/0000-0001-9578-7982; X. Wang, https://orcid.org/0000-0002-2399-6336}

\appendix
\section{}\label{appA}
We present data from related papers which observed a transition from deep convection to shallow convection as the depth ratio is altered. Each of the papers we consider uses a two-domain approach in modeling the system (so that the interface conditions can be explicitly enforced), and they each use a linear equation of state for the Boussinesq approximation. There are a variety of approaches taken in each work, from analytical methods to numerical simulations to experiments. Table \ref{tab:comparisonData1} references the approach taken in each paper, and we expand on the analytical approaches here: \cite{chen1988,mckay1998onset,straughan2002a,hirata2007linear,yin2013stability} each use linear stability theory, \cite{hill2009,mccurdy2019convection} use nonlinear stability arguments,  and \cite{han2020dynamic} investigate this bifurcation using center manifold reduction theory.

Table \ref{tab:comparisonData1} shows the marked improvements in predicting the critical depth ratio $\dhatc$ using our theory from \eqref{eq:better} compared to predictions made using the old model. Despite several works operating under different assumptions about the system-- like different governing equations (e.g., Darcy-Brinkman in lieu of Darcy), assuming the porous media has a high porosity, or using a non-Newtonian fluid-- we find our new model produces an accurate prediction for the critical depth ratio.

Our formula in \eqref{eq:better} can be rearranged to solve for the critical $\epsilon_T$ or $\Da$ values, and two works, \citet[][]{yin2013stability} and \citet[][]{han2020dynamic}, note how the transition from deep to shallow convection is affected by changing these parameters. Table \ref{tab:comparisonData2} shows our prediction for the critical values of these parameters alongside their true values. We once again find better agreement between the true values and predictions made with the new model compared to our old formula.

\begin{table}
\centering
\def\arraystretch{1.3}
\begin{tabular}{|l|l|l|l|l|l|}
\cline{1-6} 
\textbf{Paper}                                            &  \textbf{\begin{tabular}[c]{@{}l@{}}Gov. \\ Eqns\end{tabular}}                      & \textbf{Parameters}                                                                          & \textbf{\begin{tabular}[c]{@{}l@{}}``True'' critical \\ depth ratio \\ from paper\end{tabular}}               & \textbf{\begin{tabular}[c]{@{}l@{}}Predicted \\ $\dhatc$ from \\ new model\end{tabular}} & \textbf{\begin{tabular}[c]{@{}l@{}}Predicted \\ $\dhatc$ from \\ old model\end{tabular}} \\ \cline{1-6} 
\begin{tabular}[c]{@{}l@{}}\citet[][]{chen1988} \\ (analytical)\end{tabular}                               & (a) &\begin{tabular}[c]{@{}l@{}}$\sqrt{\Da}= 0.003$, \\ $\epsilon_T = 0.7$, $\alpha = 0.1$\end{tabular}                 & $\dhatc\in[0.12, 0.13]$                                                                                      & 0.1291                                                                                  & 0.1176                                                                                  \\ \cline{1-6} 
\begin{tabular}[c]{@{}l@{}}\citet[][]{chen1989experimental} \\ (experimental)\end{tabular}                            &--& \begin{tabular}[c]{@{}l@{}}$\sqrt{\Da}= 0.002$,\\ $\epsilon_T = 0.7$, $\alpha = 0.1$\end{tabular}                  & $\dhatc\in[0.1, 0.11]$                                                                                       & 0.1054                                                                                  & 0.0960                                                                                  \\ \cline{1-6} 
\begin{tabular}[c]{@{}l@{}}\citet[][]{chen1992convection}\\ (numerical)\end{tabular}                                 & (b) & \begin{tabular}[c]{@{}l@{}}$\sqrt{\Da}= 0.889\times 10^{-5}$,\\ $\epsilon_T = 0.725$\end{tabular}                  & \begin{tabular}[c]{@{}l@{}}$\dhatc\in[0.1, 0.2]$,\\ with a ``critical \\ value of $\dhatc=0.13$''\end{tabular} & 0.1287                                                                                  & 0.1172                                                                                  \\ \cline{1-6} 
\begin{tabular}[c]{@{}l@{}}\citet[][]{mckay1998onset} \\ (analytical)\end{tabular}                                   &(a) & \begin{tabular}[c]{@{}l@{}}$\Da=10^{-5}$, $\epsilon_T = 2.0$\end{tabular}                                        & $\dhatc\approx 0.2128$                                                                                       & 0.2241                                                                                  & 0.2041                                                                                  \\ \cline{1-6} 
\begin{tabular}[c]{@{}l@{}}\citet[][]{straughan2002a}\\ (analytical)\end{tabular}                                & (a) & \begin{tabular}[c]{@{}l@{}}$\sqrt{\Da} = 3.279\times 10^{-3}$, \\ $\epsilon_T = 0.7$, $\alpha = 0.78$\end{tabular} & $\dhatc\in[0.14, 0.15]$                                                                                      & 0.1350                                                                                  & 0.1229                                                                                  \\ \cline{1-6} 
\begin{tabular}[c]{@{}l@{}}\citet[][]{hirata2007linear}$^{\dagger}$\\ (analytical)\end{tabular}                                & (c) & \begin{tabular}[c]{@{}l@{}}$\sqrt{\Da} = 0.003$, \\ $\epsilon_T = 0.7$, $\alpha = 0.1$\end{tabular} & $\dhatc\in[0.12, 0.14]$                                                                                      & 0.1226                                                                                  & 0.1116                                                                                  \\ \cline{1-6}  
                                                                                  \begin{tabular}[c]{@{}l@{}}\citet[][]{hill2009}\\ (analytical)\end{tabular}                         &  (c)& \begin{tabular}[c]{@{}l@{}}$\Da = 5\times 10^{-6},$ \\ $\epsilon_T = 0.7$\end{tabular}                             & $\dhatc\in[0.115, 0.117]$                                                                                    & 0.1115                                                                                  & 0.1015                                                                                  \\ \cline{1-6} 
\multirow{2}{*}{\begin{tabular}[c]{@{}l@{}}\citet[][]{yin2013stability}\\ (analytical)\end{tabular}} & (d)         & \begin{tabular}[c]{@{}l@{}}$\sqrt{\Da} = 0.002,$ \\ $\epsilon_T=0.7,$ $\alpha=0.1$\end{tabular}                     & $\dhatc\in[0.1, 0.11]$                                                                                       & 0.1054                                                                                  & 0.0960                                                                                  \\ \cline{2-6} 
                                                                                                        & (d)& \begin{tabular}[c]{@{}l@{}}$\sqrt{\Da} = 0.002,$\\  $\epsilon_T=0.6,$ $\alpha=0.1$\end{tabular}                    & $\dhatc\in[0.09, 0.10]$                                                                                      & 0.0976                                                                                  & 0.0889                                                                                  \\ \cline{1-6} 
\multirow{4}{*}{\begin{tabular}[c]{@{}l@{}}\citet[][]{mccurdy2019convection} \\ (analytical)\end{tabular}} & (a)&  \begin{tabular}[c]{@{}l@{}}$\sqrt{\Da} = 5\times 10^{-3},$ \\ $\epsilon_T = 0.7$, $\alpha = 1.0$\end{tabular}      & $\dhatc\approx .181$                                                                                         & 0.1667                                                                                  & 0.1518                                                                                  \\ \cline{2-6} 
                                                                                                      &   (a)& \begin{tabular}[c]{@{}l@{}}$\sqrt{\Da} = 1\times 10^{-3},$\\  $\epsilon_T = 0.7$, $\alpha = 1.0$\end{tabular}      & $\dhatc\approx 0.079$                                                                                        & 0.0745                                                                                  & 0.0679                                                                                  \\ \cline{2-6} 
                                                                                                  &      (a) & \begin{tabular}[c]{@{}l@{}}$\sqrt{\Da} = 5\times 10^{-3},$\\  $\epsilon_T = 0.5$, $\alpha = 1.0$\end{tabular}      & $\dhatc\approx 0.115$                                                                                        & 0.1409                                                                                  & 0.1283                                                                                  \\ \cline{2-6} 
                                                                                                        &(a) & \begin{tabular}[c]{@{}l@{}}$\sqrt{\Da} = 5\times 10^{-3},$\\ $\epsilon_T = 1.0$, $\alpha = 1.0$\end{tabular}       & $\dhatc\approx 0.256$                                                                                        & 0.2440                                                                                  & 0.2222                                                                                  \\ \cline{1-6} 
\begin{tabular}[c]{@{}l@{}}\citet[][]{han2020dynamic}\\ (analytical)\end{tabular}  &(a)                         & \begin{tabular}[c]{@{}l@{}}$\Da = 25\times 10^{-6},$\\ $\epsilon_T = 0.7$, $\alpha = 1.0$\end{tabular}             & $\dhatc\in[0.16, 0.17]$                                                                                      & 0.1667                                                                                  & 0.1518                                                                                                                                                    \\ \cline{1-6} 
\end{tabular}
  \caption{Data from papers which note a critical depth ratio along with the prediction for the $\dhatc$ from our theory \eqref{eq:better} compared to the prediction made using the old model. The approach taken by the authors is noted alongside the citation, with a majority of the analytic approaches taking the form of investigating marginal stability curves and \citet[][]{han2020dynamic} making use of center manifold reduction theory to determine where the transition occurs. 
  Notation for the governing equations: (a) Navier-Stokes-Darcy-Boussinesq; (b) Navier-Stokes-Darcy-Brinkman-Boussinesq, with nonlinear Forchheimer term in Darcy; (c) Navier-Stokes-Darcy-Brinkman-Boussinesq;  (d) Navier-Stokes-Darcy-Boussinesq with an Oldroyd-B fluid. $\dagger$ from the two-domain approach in \citet[][]{hirata2007linear}.  }
  \label{tab:comparisonData1}
\end{table}

\begin{table}
\centering
\def\arraystretch{1.3}
\begin{tabular}{ll|l|l|l|l|}
\cline{1-6} 
\multicolumn{1}{|l|}{\textbf{Paper}}                                            &  \textbf{\begin{tabular}[c]{@{}l@{}}Gov. \\ Eqns\end{tabular}}                      & \textbf{Parameters}                                                                          & \textbf{\begin{tabular}[c]{@{}l@{}}``True'' critical \\ $\epsilon_T$ value \\ from paper\end{tabular}}               & \textbf{\begin{tabular}[c]{@{}l@{}}Predicted \\ $\epsilon_T^*$ from \\ new model\end{tabular}} & \textbf{\begin{tabular}[c]{@{}l@{}}Predicted \\ $\epsilon_T^*$ from \\ old model\end{tabular}} \\ \cline{1-6} 
\multicolumn{1}{|l|}{\begin{tabular}[c]{@{}l@{}}\citet[][]{yin2013stability}\\ (analytical)\end{tabular} }& (i)         & \begin{tabular}[c]{@{}l@{}}$\sqrt{\Da} = 0.002,$ \\ $\hat{d}=0.1,$ $\alpha=0.1$\end{tabular}                     & $\epsilon_T^*\in[0.6, 0.7]$                                                                                       & 0.6298                                                                                  & 0.7595                                                                                  \\ \cline{1-6}                                                                   
\multicolumn{1}{|l|}{\begin{tabular}[c]{@{}l@{}}\citet[][]{han2020dynamic}\\ (analytical)\end{tabular}}  &(ii)                         & \begin{tabular}[c]{@{}l@{}}$\Da = 25\times 10^{-6},$\\ $\hat{d} = 0.2$, $\alpha = 1.0$\end{tabular}             & $\epsilon_T^*\in[0.8, 0.9]$                                                                                      & 1.0078                                                                                  & 1.2151                                                                                                                                                    \\ \cline{1-6}\\ \cline{1-6} 
\multicolumn{1}{|l|}{\textbf{Paper}     }                                       &  \textbf{\begin{tabular}[c]{@{}l@{}}Gov. \\ Eqns\end{tabular}}                      & \textbf{Parameters}                                                                          & \textbf{\begin{tabular}[c]{@{}l@{}}``True'' critical \\ $\Da$ value \\ from paper\end{tabular}}               & \textbf{\begin{tabular}[c]{@{}l@{}}Predicted \\ $\Da^*$ from \\ new model\end{tabular}} & \textbf{\begin{tabular}[c]{@{}l@{}}Predicted \\ $\Da^*$ from \\ old model\end{tabular}} \\ \cline{1-6} 
\multicolumn{1}{|l|}{\begin{tabular}[c]{@{}l@{}}\citet[][]{yin2013stability}\\ (analytical)\end{tabular}} & (i)         & \begin{tabular}[c]{@{}l@{}}$\epsilon_T = 0.7,$ \\ $\hat{d}=0.1,$ $\alpha=0.1$\end{tabular}                     & $\Da^*\approx 1.6\times10^{-5}$                                                                                       & $1.6395\times10^{-5} $                                                                                 & $2.3836\times10^{-5}$                                                                                  \\ \cline{1-6}                                                                   
\multicolumn{1}{|l|}{\begin{tabular}[c]{@{}l@{}}\citet[][]{han2020dynamic}\\ (analytical)\end{tabular}}  &(ii)                         & \begin{tabular}[c]{@{}l@{}}$\epsilon_T = 0.7,$\\ $\hat{d} = 0.2$, $\alpha = 1.0$\end{tabular}             & $\Da^*\in\left(7\times 10^{-4}, 8\times 10^{-4}\right)$                                                                                      & $5.1815\times 10^{-5}$                                                                                  & $7.5333\times 10^{-5}$                                                                                                                                                   \\ \cline{1-6} 
\end{tabular}
  \caption{Data from related papers which observed a transition from deep convection to shallow convection as the ratio of thermal diffusivities $\epsilon_T$ or the Darcy number $\Da$ is altered. We present our prediction for the critical value from the theory presented in \eqref{eq:better} (rearranged to solve for $\epsilon_T^*$ or $\Da^*$ instead of $\dhatc$) compared to the prediction made using the old model. The approach taken by the authors is noted alongside the citation, with \citet[][]{yin2013stability} investigating marginal stability curves and \citet[][]{han2020dynamic} making use of center manifold reduction theory to determine where the transition occurs. 
  Notation for the governing equations: (i) Navier-Stokes-Darcy-Boussinesq with an Oldroyd-B fluid; (ii) Navier-Stokes-Darcy-Boussinesq. }
  \label{tab:comparisonData2}
\end{table}

\bibliographystyle{jfm}
\bibliography{theory_paper}

\end{document}